\documentclass[a4paper,10pt]{article}

\usepackage[a4paper, total={6.5in, 8.5in}]{geometry}
\usepackage[utf8]{inputenc}
 \usepackage{graphics}
 \usepackage{graphicx}
 \usepackage{setspace}

\usepackage{amssymb}
\usepackage{amsmath}
\usepackage{amsthm}

\title{The role of magnetic excitations in magnetoresistance  \\
 and Hall effect of
slightly  TM-substituted BaFe$_{2}$As$_2$ \\
 compounds (TM = Mn, Cu, Ni)}
\author{J. P. Pe\~na$^1$, M. M. Piva$^2$, C. B. R. Jesus$^2$, G. G. Lesseux$^2$, T. M. Garitezi$^2$, D. Tobia$^2$, \\
P. F. S. Rosa$^{2,3}$, T. Grant$^3$, Z. Fisk$^3$,
C. Adriano$^2$, R. R. Urbano$^2$, P. G. Pagliuso$^2$, P. Pureur$^1$}

\begin{document}

\maketitle

\begin{center}
 $^1$Instituto de F\'isica, Universidade Federal do Rio Grande do
Sul, Av. Bento Gon\c{c}alves 9500, C.P. 15051, 91501-970, 
Porto Alegre, RS, Brazil\\

$^2$Instituto de F\'isica Gleb Wataghin, Universidade Estadual de Campinas, Rua Sérgio Buarque de
Holanda 777, C.P. 13083-970, 
Campinas, SP, Brazil\\

$^3$Department of physics and Astronomy, University of California, 2186 Frederick Reines Hall, Irvine, CA, C.P 92697-4574, 
USA
\end{center}

\begin{abstract}

We report on electrical resistivity, magnetoresistance (MR) and Hall effect measurements in four non-superconducting
BaFe$_{2-x}$TM$_x$As$_2$ (TM = Mn, Cu and Ni) single crystals with small values of the chemical substitution $x$. 
The spin density wave (SDW) ordering that occurs in these systems at temperatures $T\sim$ (120  – 140) K, in close
vicinity to a tetragonal/orthorhombic transition, produces 
significant modifications in their magneto-transport properties. While in the magnetically ordered phase 
the MR is positive and its magnitude increases with decreasing temperatures, in the paramagnetic regime
the MR becomes vanishingly small. Above the spin density wave transition temperature
($T_{\text{SDW}}$) the Hall coefficient $R_H$  is negative, small and weakly temperature 
dependent, but a remarkable change of slope occurs in the $R_H$ versus $T$ curves at $T = T_{\text{SDW}}$.
The Hall coefficient amplitude, while remaining negative, increases steadily and 
significantly as the temperature is decreased below $T_{\text{SDW}}$ and down to $T =$ 20 K. The qualitative behavior 
of both MR and Hall coefficient is weakly dependent on the chemical substitution in the studied limit. The experiments
provide strong evidence that scattering of charge carriers by magnetic excitations has to be taken into
account to explain the behavior of the resistivity, magnetoresistance and Hall effect in the ordered 
phase of the studied compounds. Effects of multiple band conduction also must be considered for a complete interpretation of the results.

\end{abstract}


%\doublespacing
\section{Introduction}

Because of its peculiar electronic properties along with its similarities with the high temperature 
cuprate superconductors (HTCS), the Fe-pnictide superconductors have fascinated scientists since its discovery 
in 2008 \cite{hosono}. 
As the cuprates, the Fe-pnictide superconductors have parent compounds
showing an antiferromagnetic ground state. Also similar to the HTCS, superconductivity in the Fe-pnictides
can be achieved by doping the precursor compounds with 
electrons or holes. %; however, there is still a debate whether the superconducting state in the Fe-pnictides emerges
%due to effective doping or chemical substitution.
Contrasting with the cuprates,
superconductivity in the Fe-pnictides may also be obtained  by applying pressure \cite{paglione}.
Further, the precursors of the iron based superconductors 
are metallic and the antiferromagnetism is related to stabilization of a spin density wave (SDW)
state \cite{singh, dong}. Nevertheless, it is currently accepted that the simultaneous presence 
of both localized and itinerant moments is necessary to explain magnetism, transport and other 
electronic properties of these compounds \cite{harriger, zhao, liu}.
The typical $T-x$ phase diagram of Fe-pnictides shows that the parent compounds and underdoped
samples exhibit  a tetragonal, paramagnetic phase at high temperatures, while an orthorhombic, antiferromagnetic 
phase characterizes the low-temperature state. The structural and magnetic
transitions are gradually suppressed upon doping the parent compound with selected impurities,
and above a certain threshold of the chemical
substitution, a superconducting ground state is stabilized \cite{paglione}.

Among several known families of iron based superconductors, the 122 family is the most studied.  
One of the parent compounds of this family is the BaFe$_2$As$_2$,
which displays a structural (tetragonal to orthorhombic) transition closely followed by a SDW phase transition 
at \mbox{$T\sim$ 140 K}. 
Furthermore, a  superconducting transition can be induced by external pressure or chemical substitutions.
The substitutions can be either out or inside
the Fe-As planes. The first case is represented by the partial substitution of Ba atoms by K atoms (hole doping). Substitution 
inside the planes are achieved by partially substituting the Fe atoms by
Co, Ni, Cu, Rh or Pd  \cite{paglione, ishida, piva, garitezi, canfield, ni} (electron
doping), or the As atoms by P atoms (isovalent substitution).
All these substitutions disrupt the magnetic order and, depending on the impurity concentration, induce superconductivity.
However as  similar results are obtained by applying external pressure, 
it is still controversial whether  chemical substitution in fact leads to an effective charge 
doping and is the main factor driving superconductivity in the BaFe$_2$As$_2$ compound \cite{yamazaki, rosa}.
On the other hand, although substitutions such as
Mn \cite{kim} and Cr \cite{sefat} also suppress the SDW phase, the superconducting state is not 
observed due to the strong magnetic pair-breaking mechanism induced by these impurities.

An important feature in the electronic structure of the Fe-pnictides is the presence of both electron and
hole pockets on their Fermi surface \cite{paglione, ishida}. As a consequence,
it has been difficult to identify the microscopic phenomena giving place to magnetism 
and many doubts remain on the mechanisms governing the  electric charge transport in these systems.
To shed new light on the discussion of this specific subject,  we have studied  electrical magneto-transport properties
in several non-superconducting compounds of the Ba-122 family. In particular, 
we have experimentally investigated the magnetoresistance (MR) and Hall effect (HE) of 
BaFe$_{2-x}$TM$_x$As$_2$  (TM = Mn, Cu, Ni) single crystal samples in the very low doping limit. We also report on resistivity
versus temperature measurements at zero applied magnetic field.

In general, the transversal MR ($\mu_0\text{H}\parallel c\perp I$, where $I$ represents the electrical current and $c$ is the principal
symmetry axis)
of 122 compounds is positive and
displays an unusual  temperature dependence becoming
negligibly small for \mbox{$T > T_{\text{SDW}}$} \cite{kuo, huynh}. On the other hand, the longitudinal
MR ($\mu_0\text{H}\parallel ab\parallel I$ where $ab$ refers to the planar orientation)
is negative, and tough small, it is still measurable in
the paramagnetic region \cite{albenque2}. The longitudinal MR behavior in the disordered region was explained as
resulting from the spin-disorder suppression mechanism produced by the external magnetic field \cite{albenque2}.
Conversely, no explanation is found in the literature for the temperature and field dependent behavior 
of the transversal MR, which is the one studied here.  The Hall coefficient ($R_H$)
shows a strong temperature dependence in compounds of the 122 family \cite{albenque, fang}. In particular, the absolute
value of $R_H$ has a drastic increase when the temperature decreases below  $T_{\text{SDW}}$ in  non-doped and slightly doped
compounds. Some attempts to  explain this behavior include (i) a Fermi surface
reconstruction at $T_{\text{SDW}}$  \cite{albenque, fang, yi}, or (ii) the effect of conduction by multiple bands.
However, in undoped and slightly doped compounds as those studied here, the magnetic
transition is very sharp, in some cases considered weakly first order \cite{chu}, 
so that the first  interpretation does not adequately explain the observed
variations of $R_H$  in temperatures far below $T_{\text{SDW}}$. 
The second explanation implies that, if not accompanied by another effect, strong modifications  in the electronic
band structure should occur all along  the magnetic ordered phase. 
% In addition, $R_H$ has a marked signal in the whole regime below the room temperature. This is true 
% even for the non-doped compound which is almost completely compensated. In the only frame of the Boltzmann distribution, this 
% fact must be explained by some difference in the mobilities of holes and electrons doing one  dominant over the other.
% However, the mechanism that generates such behavior, even in the paramagnetic phase is unknown.
These controversies show that the mechanisms governing the magneto-transport phenomena in the 122 Fe-pnictides
are still not completely understood.
In this work we present evidences that carrier scattering by magnetic excitations leading to anomalous contributions both
to the MR and HE should be considered as a relevant mechanism for describing the magneto-transport properties of these systems

The results here described strongly suggest that not only multiple-band conduction, but
also scattering by  magnetic excitations, must be taken into account for
explaining both the MR and HE in the magnetically ordered ground state of 
the undoped and slightly doped 122 Fe-pnictides.

\section{Experimental details}

Single crystals of BaFe$_{2-x}$TM$_x$As$_2$, $x=0$ and $x=0.020$, 0.012 and 0.015 were synthesized 
for TM = Mn, Cu and Ni, respectively.
These concentrations were estimated by Energy Dispersive Spectroscopy (EDS) and 
wavelength-dispersive X-ray spectroscopy (WDS) analyses. We further compared the measured SDW temperatures
with predictions of the phase diagrams found in literature. 
This last criterion is useful because of the large uncertainties 
related to the employed EDS and WDS techniques in the limit of low impurity concentration.
The crystals were grown by using the In-flux method as reported
in \mbox{Ref. \cite{garitezi}}. None of our four samples shows superconductivity and no detwinning 
processes were attempted on the obtained crystals.

X-rays diffraction and resistivity measurements were performed for characterization. 
The magnetic ordering temperature $T_{\text{SDW}}$ was estimated
from the temperature derivative of the zero field electrical resistivity curves.
Table \ref{parameters} displays the values of $T_{\text{SDW}}$ and  the $c-$axis lattice
parameters extracted from the experiments above mentioned.

\begin{table}[h]
\centering
\caption{Lattice parameter along the $c-$axis  and the magnetic transition 
temperatures for the studied samples of BaFe$_{2-x}$TM$_x$As$_2$ (TM = Mn, Cu, Ni).}\label{parameters}
\begin{tabular}{|c|c|c|c|}\hline
\textbf{Sample} & $c$ (\AA) &  $T_{\text{SDW}}$ (K) \\%$T_{\text{SDW}}$\pm1$ (K) &
\hline 
$x=0$ &  13.021 $\pm0.001$ & 135 $\pm3$\\% 129.5 &    
\hline 
Mn$_{0.020\pm0.006}$ & 13.022 $\pm0.003$ & 121 $\pm3$\\% 115.1 &
\hline 
Cu$_{0.012\pm0.004}$ & 13.011 $\pm0.001$ & 123 $\pm1$\\%121.4 &
\hline 
Ni$_{0.015\pm0.005}$ & 13.034$\pm0.002$  & 121 $\pm2$\\%116.2 & 
\hline 
\end{tabular}
\end{table}

Electric transport measurements were carried out by using a standard four probe method. The electric contacts 
were made with silver epoxy and Cu wires in platelet-like crystals. The contact leads for voltage measurements
were attached to the same edge of the sample for measuring the longitudinal resistance, and
to opposite edges for the transversal resistance. Measurements were performed upon the application of  magnetic fields 
with magnitude between 0 and $\pm$9 T and orientation parallel to the $c-$axis. The magnetoresistance was determined from 
the average $\rho_{even}=(\rho_++\rho_-)/2$ of the longitudinal measurement and the Hall resistivity from 
$\rho_{odd}=(\rho_+-\rho_-)/2$ of the transversal voltage measurement.
The term $\rho_{+/-}$ refers to the vertical direction positive/negative of the magnetic field, respectively. 
All  the magneto-transport experiments were performed with
a low-frequency AC bridge of a commercial PPMS@ platform manufactured by Quantum Design, Inc.

\section{Results and discussion}

\subsection{Resistivity}

\begin{figure}
 \centering
\includegraphics[keepaspectratio,width =7.2truecm]{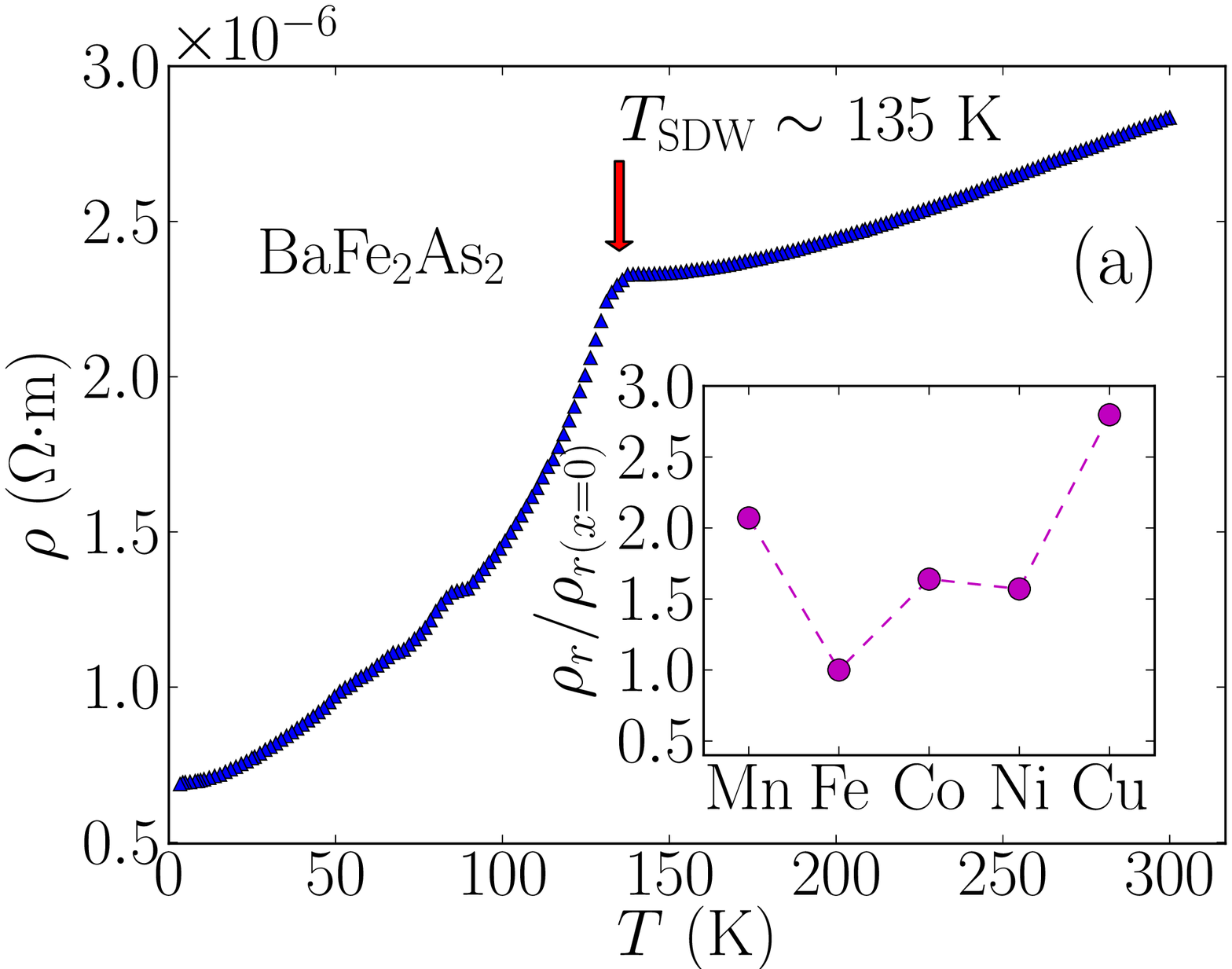}\\
\includegraphics[keepaspectratio,width =7.2truecm]{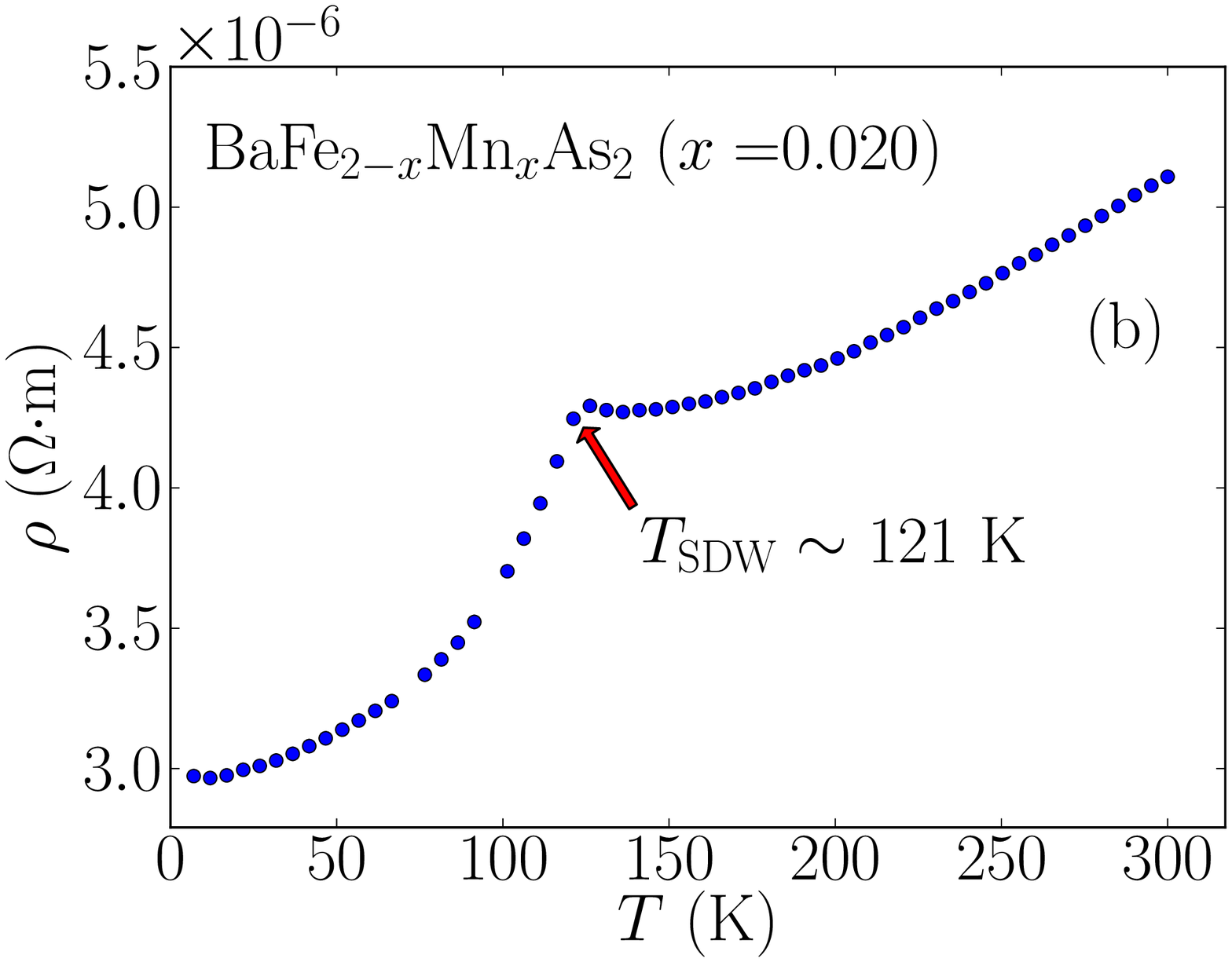}\\
\includegraphics[keepaspectratio,width =7.2truecm]{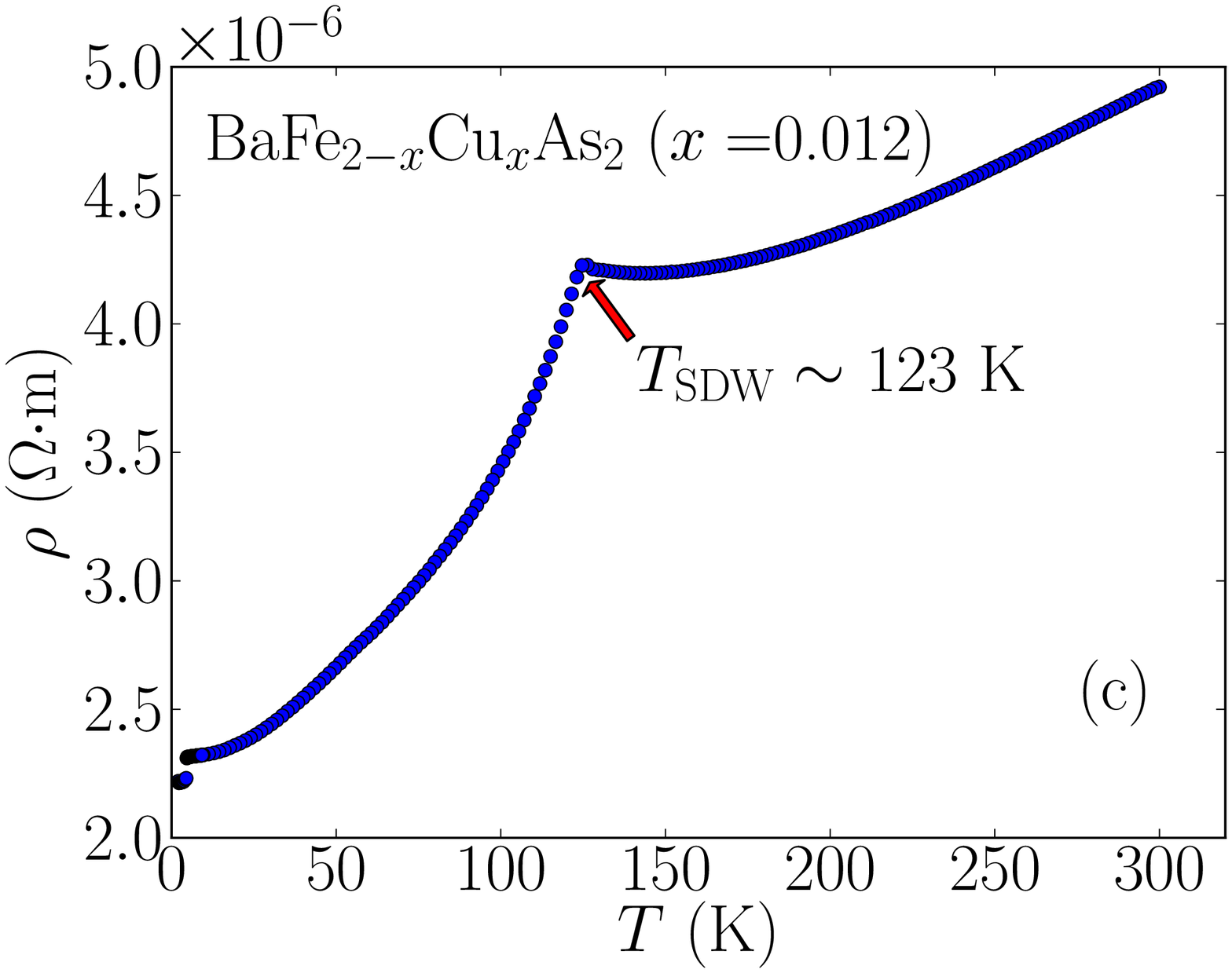}\\
\includegraphics[keepaspectratio,width =7.2truecm]{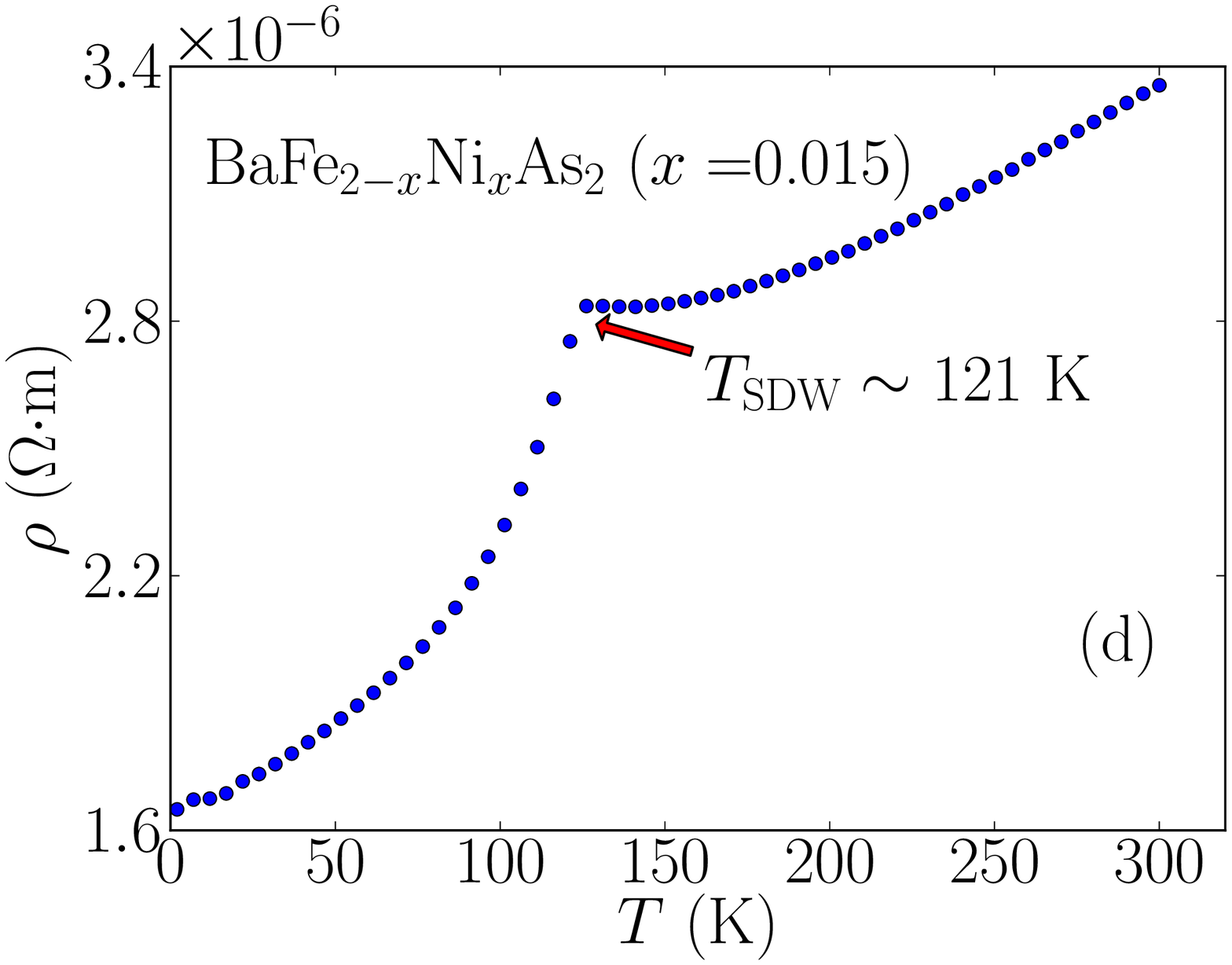}
\caption{Resistivity as a function of the temperature for the four studied samples. The magnetic ordering 
temperatures ($T_{\text{SDW}}$) are indicated. The inset in the (a) panel shows the ratio between the residual 
resistivity of the substituted-samples with respect to that of the pure compound (see text). } \label{RT}
\end{figure}

The  temperature-dependent electrical resistivity curves for the  studied samples are 
presented in \mbox{Fig. \ref{RT}}. The magnetic transition is  signaled by a typical kink in the \mbox{$\rho$ vs. $T$} curves.
The transition  temperature is clearly diminished upon chemical substitution.
In addition, the absolute values of the resistivity in the substituted compounds are
significantly larger than that in the parent compound.
 Thus in the 
low $x$ limit, the studied chemical substitutions seem to play the role of scattering centers increasing the relaxation 
rate rather than as donor/receptor atoms increasing the electron/holes
density. %This point will be retaken in the Hall effect analyses in section 3.3.
The inset in the (a) panel of Fig. \ref{RT} shows the residual resistivities ($\rho_r$) of the substituted samples compared 
with that of the pure compound; there we include a point obtained 
from Ref. \cite{olariu} for a sample of BaFe$_{1.98}$Co$_{0.02}$As$_2$.
The data for the substituted samples are calculated for $x=0.01$, then normalized with respect 
to the residual resistivity of the pure specimen.  
We note that the  behavior of $\rho_r$ across the series of compounds is similar to that observed in 
dilute alloys, where $\rho_r$ increases when the charge of the impurity relative to that of the host also increases
because of its higher cross section.

A close look at the $\rho$ vs. $T$ curves in Fig. \ref{RT} near the kink observed at $T_{\text{SDW}}$  reveals that the resistivity 
goes through a maximum in the form of a faint cusp at this temperature. Although the magnetic transition
is accompanied by a slight orthorhombic distortion in the studied compounds  \cite{paglione}, we rather ascribe this maximum 
to the opening of a small gap, or pseudogap, near the Fermi level because of the antiferromagnetic ordering. This maximum 
is often identified as the super-zone effect, commonly observed in the resistivity of antiferromagnetic 
metals near the ordering temperature \cite{meaden}. 
Alternatively, it has been argued that, in the absence of magnetic field, the kink in
the $\rho \text{ vs. } T$ plot, where the resistivity
drops sharply below $T_{\text{SDW}}$, can be explained by the existence of one or two (one for each type
of carrier) successive Lifshitz transitions \cite{wang}.

% More detais on superzone effect in R.J. Elliot and F.A. Wedgwood, Proc. Phys. Soc 81, 846 (1963) and Phys. Soc 84, 63 (1964) 

\subsection{Magnetoresistance}

\begin{figure}
 \centering
 \includegraphics[keepaspectratio,width =7truecm]{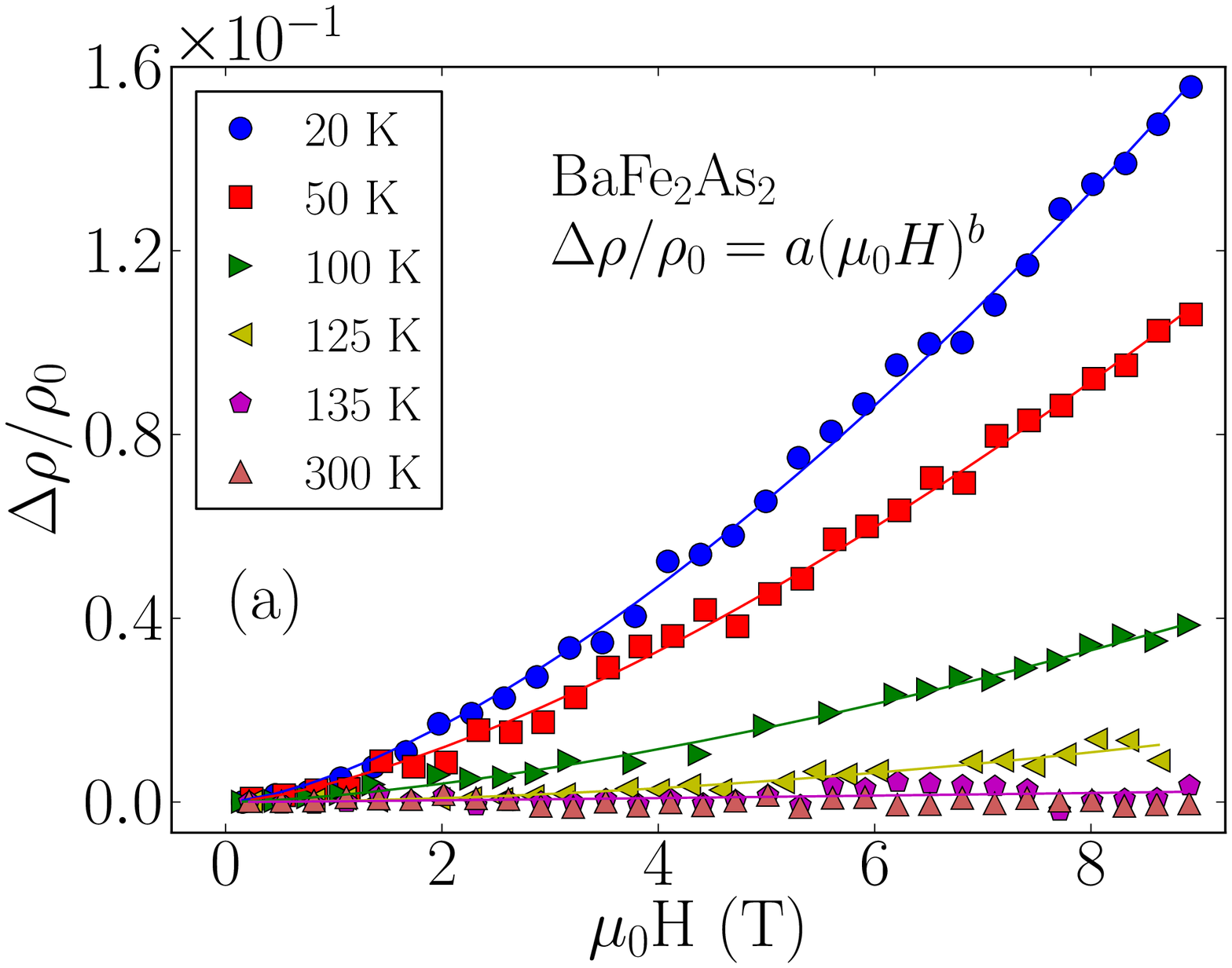} 
 \includegraphics[keepaspectratio,width =7truecm]{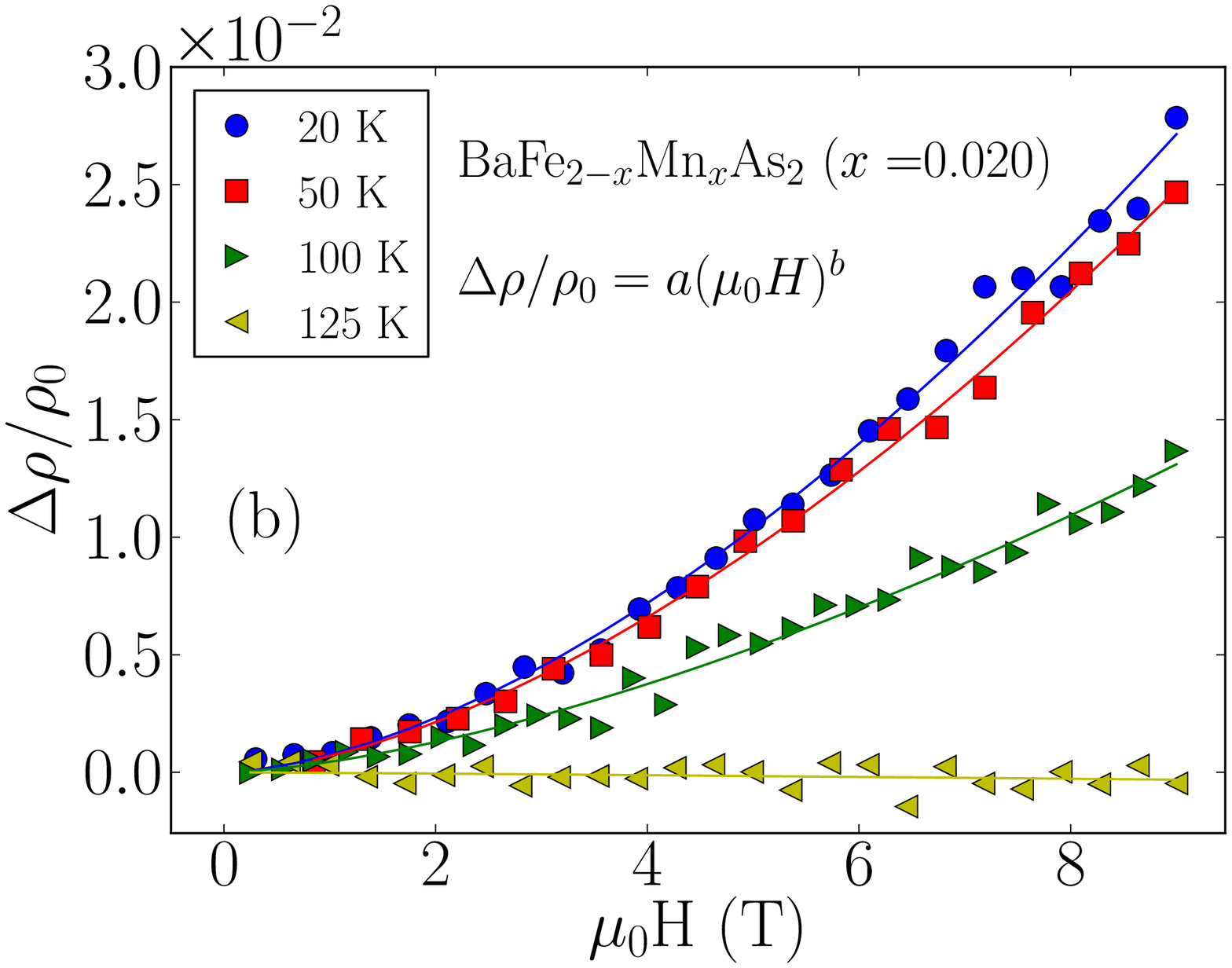} 
 \includegraphics[keepaspectratio,width =7truecm]{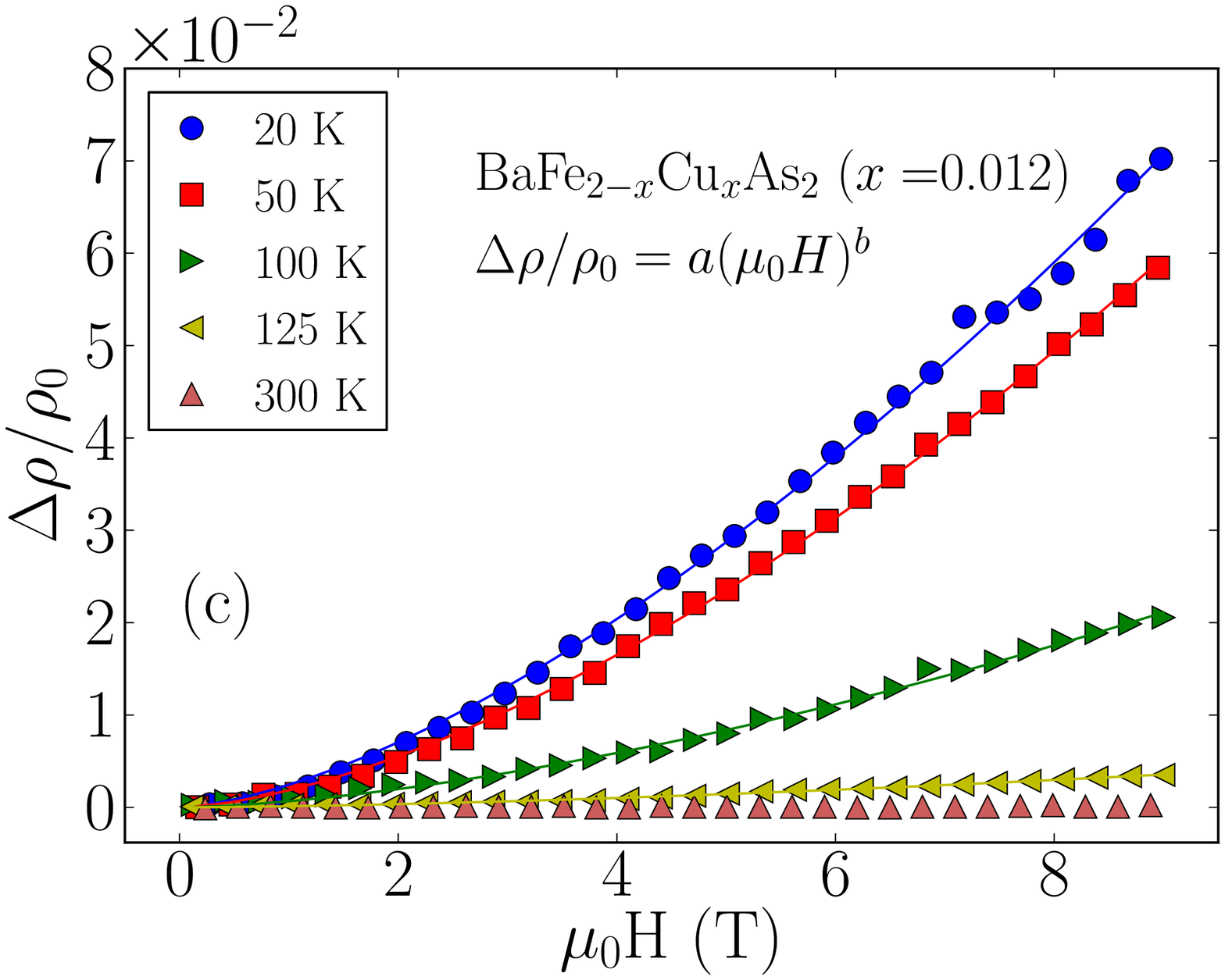} 
 \includegraphics[keepaspectratio,width =7truecm]{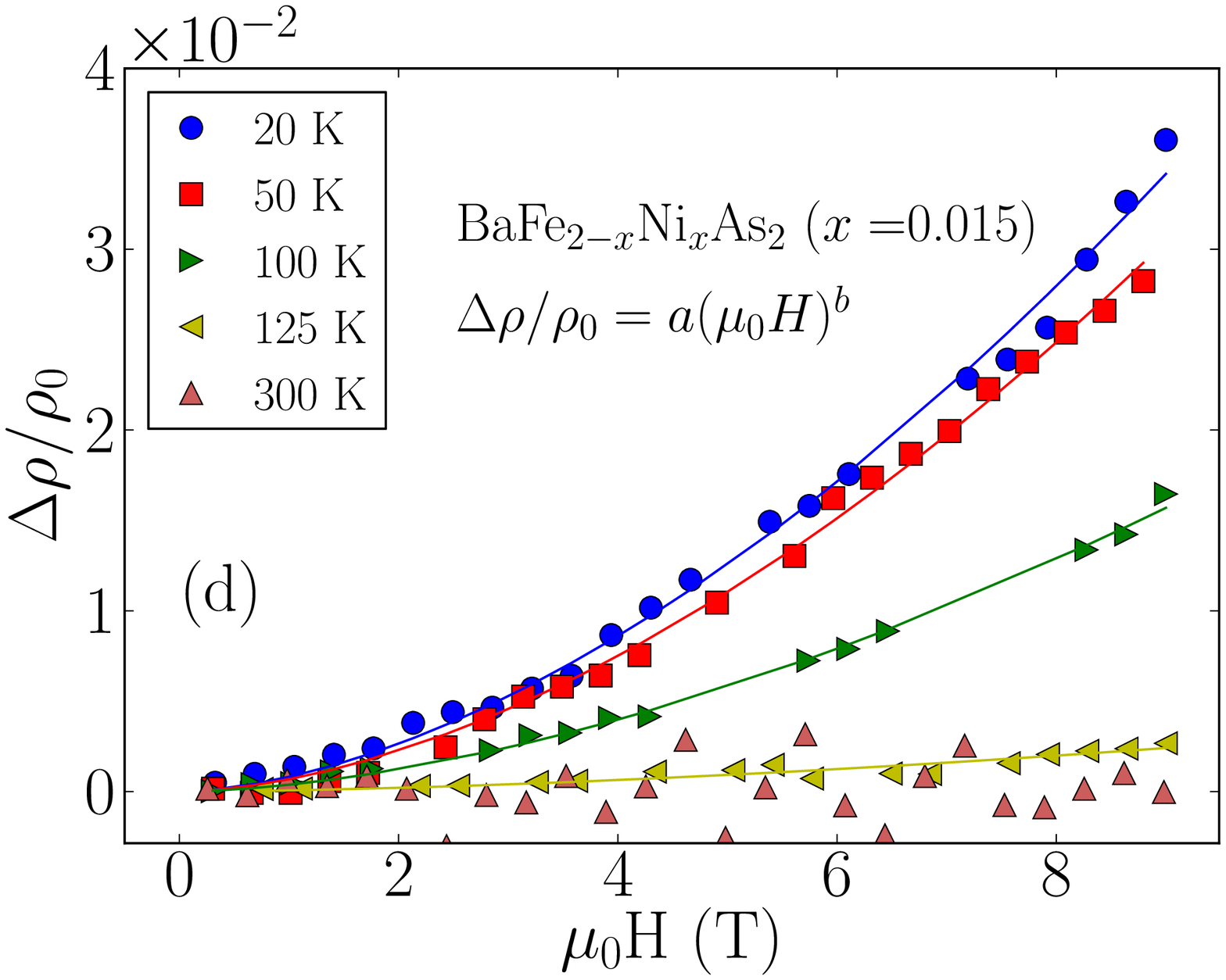} 
      \caption{MR as a function of the magnetic field ($\mu_0$H$\parallel c$) in several fixed temperatures.
     Solid lines are fits to  $\Delta\rho/\rho_0=a(\mu_0\text{H})^b$, where $b\approx1.5$ for all samples
     and temperatures. The MR is nearly zero for temperatures above $T_{\text{SDW}}$.} \label{MR_H}
\end {figure}

\begin{figure}
 \centering
\includegraphics[keepaspectratio,width =7truecm]{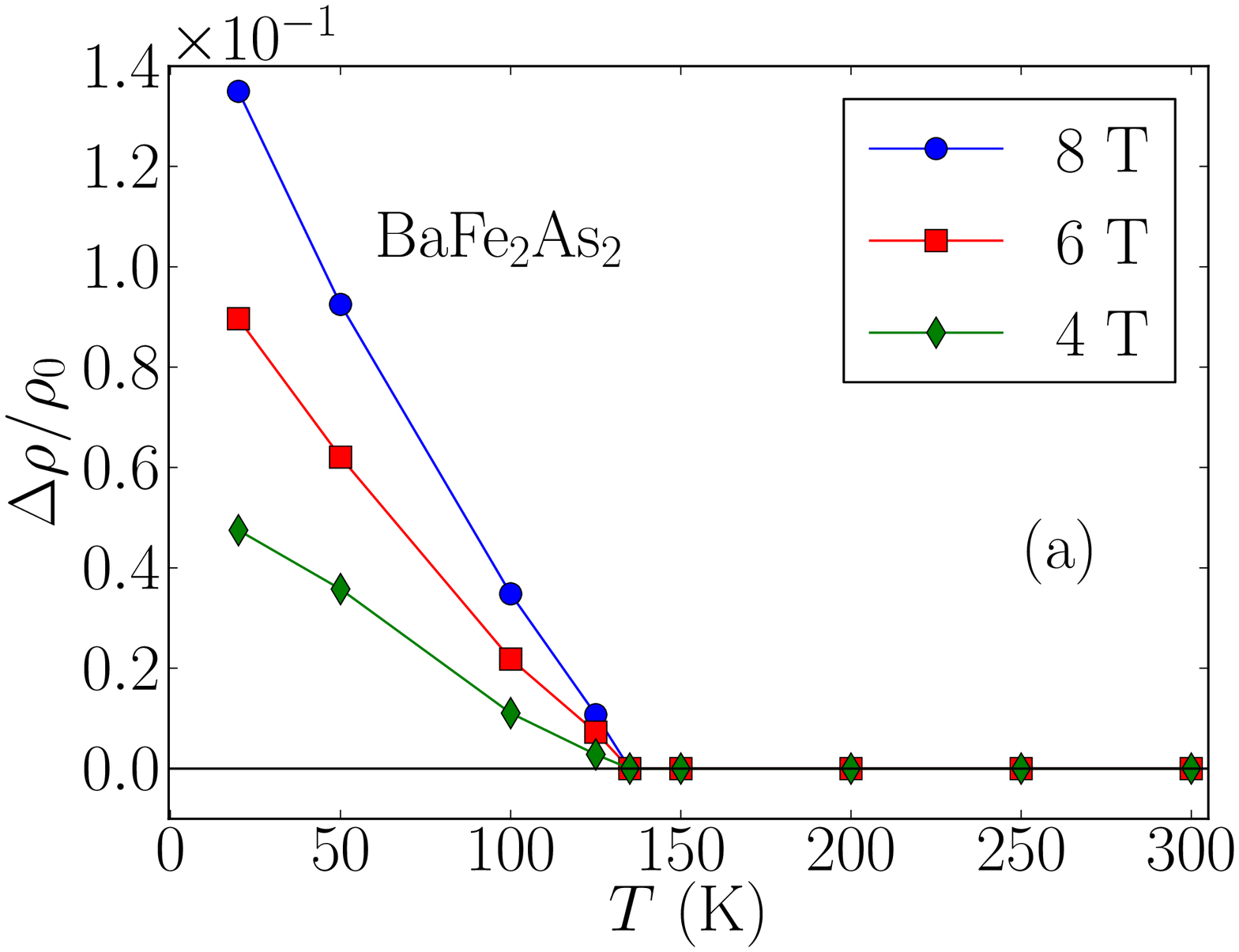}
   \includegraphics[keepaspectratio,width =7truecm]{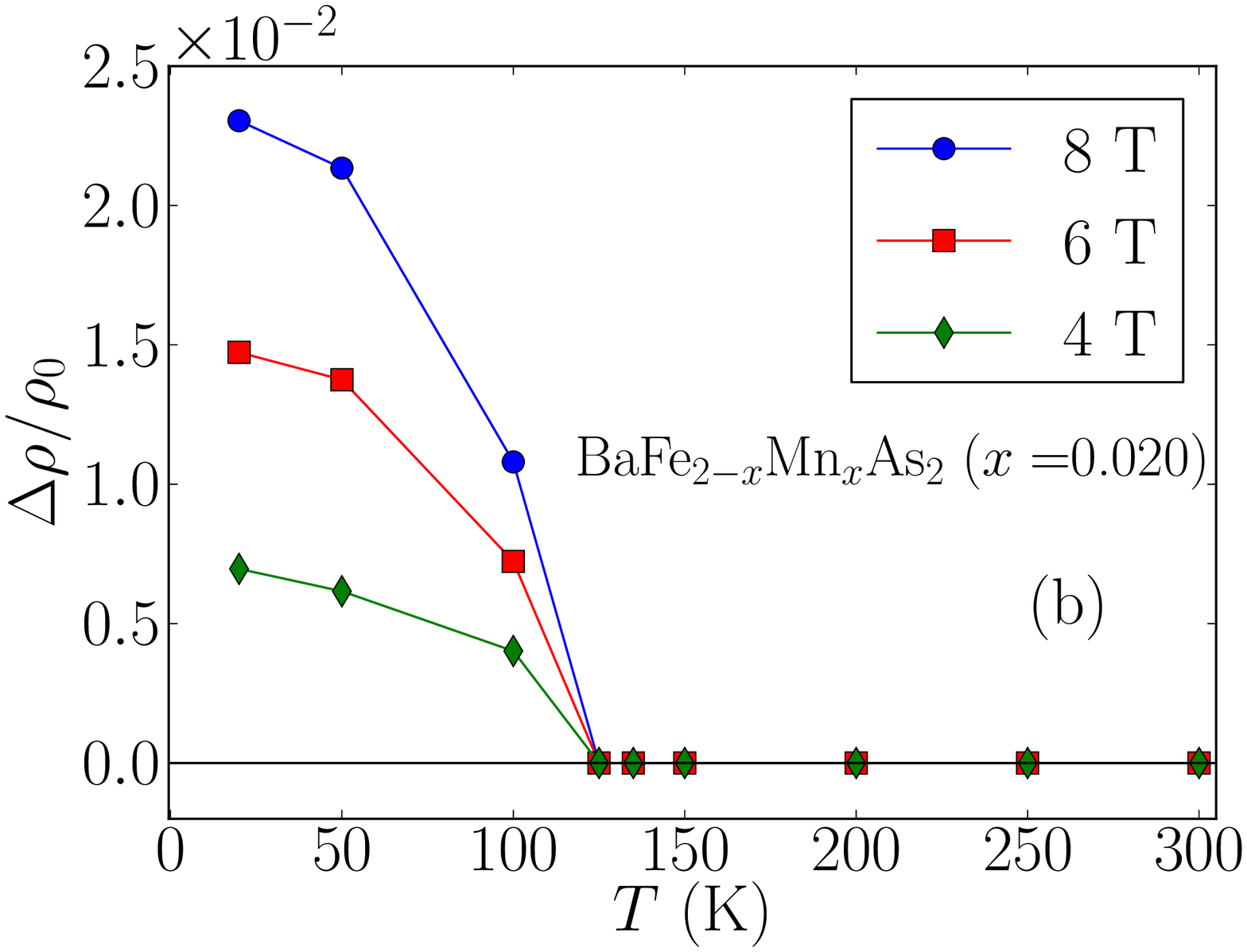} 
  \includegraphics[keepaspectratio,width =7truecm]{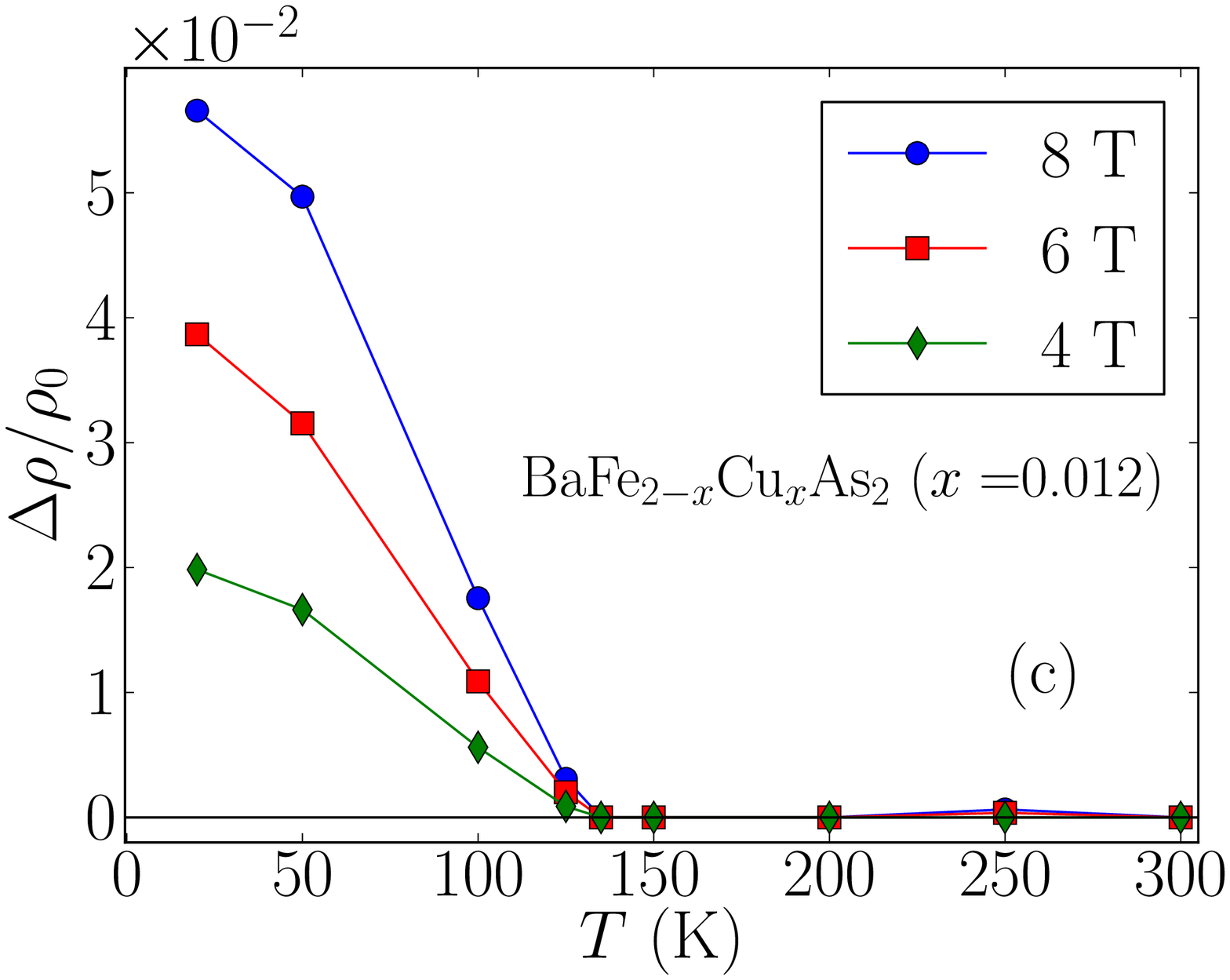}
   \includegraphics[keepaspectratio,width =7truecm]{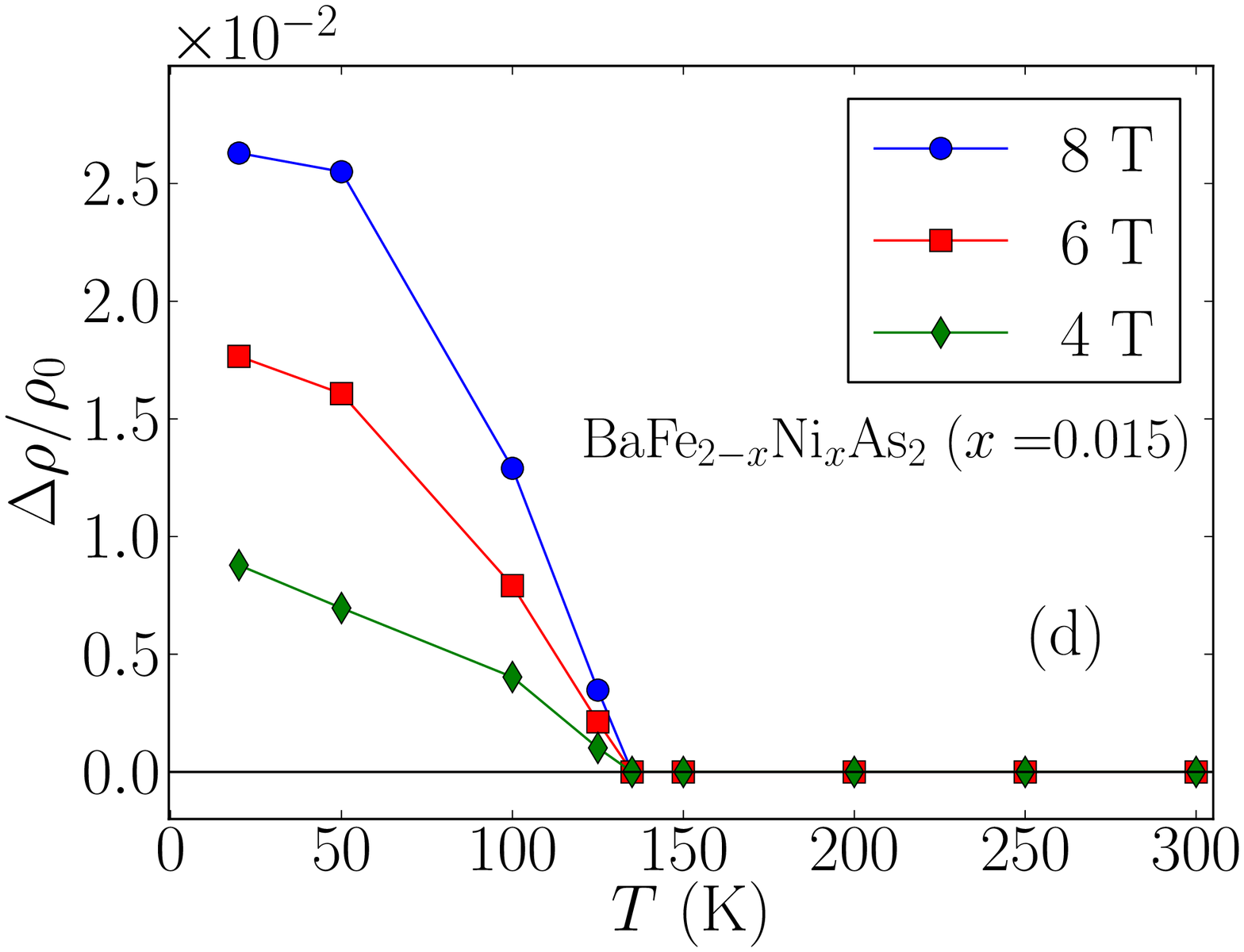}
      \caption{MR amplitude as a function of the temperature at three different fields.} \label{MR_T}
\end {figure}

In Fig. \ref{MR_H}, the MR is shown as a function of the magnetic
field at several fixed temperatures for all the studied samples. In \mbox{Fig. \ref{MR_T}} the temperature dependence of the 
MR is presented at the fixed fields $\mu_0$H = 4, 6 and 8 T. In both figures
the MR is given as $\Delta\rho/\rho(0)$ where $\Delta\rho=\rho(\mbox{H})-\rho(0)$.

Figure \ref{MR_T} shows that the MR amplitude is negligible in temperatures
above $T_{\text{SDW}}$ for all studied samples. This particular result strongly suggests that the electrical 
transport in these systems is not a single-band conduction process. 
According to the simplest two-band model, 
the low-field MR is given as \cite{ziman}:
\begin{equation}\label{eqMR}
 \Delta\rho\approx\frac{\sigma_h\sigma_e(\mu_h-\mu_e)^2}{(\sigma_h+\sigma_e)^2}\text{H}^2,
\end{equation}
 where $\sigma_{h(e)}$ is the  conductivity of the hole (electron) band and $\mu_{h(e)}$ is the respective mobility. 
 In view of the results in \mbox{Fig. \ref{MR_T}}, the  above expression ensures that the mobilities 
 for holes and electrons are approximately the same above $T_{\text{SDW}}$.
The situation is quite distinct in the ordered phase, where the MR is fairly large and positive. 
% Authors in Ref. \cite{ishida2} showed that at least three carrier types contribute to the charge transport 
% in the ordered phase  of BaFe$_2$As$_2$; thus,
In this region Eq. (\ref{eqMR}) does not describe adequately the observed results.
At first, as shown in Fig. \ref{MR_H},
the experimentally determined MR may be described as $\Delta\rho\propto a\text{H}^b$, where $a$ is a constant and  
 $b\approx$ 1.5 for all samples and temperatures. This behavior deviates 
from the quadratic field dependence predicted by Eq. (\ref{eqMR}).
Secondly,  within the two-bands scenario one would have to suppose that  some severe
Fermi surface reconstruction at $T = T_{\text{SDW}}$ \cite{albenque, yi} forces the mobilities
$\mu_e$ and  $\mu_h$ to become significantly different at this temperature. Moreover, this difference should
increase as the temperature decreases into 
 the magnetically ordered state.

We propose a description of the MR data in the temperature range below $T_{\text{SDW}}$ by considering the effect  of 
spin-dependent scattering. Since the 122 Fe-pnictides order antiferromagnetically, 
the application of a magnetic field disrupts the imbalance of the staggered magnetizations. Then, 
instead of reducing spin-disorder, as is the case in ferromagnetic metals, the field enhances 
the cross-section for spin-flip scattering. 
Roughly, one would expect that  the field-dependent resistivity 
increases as \cite{kasuya}:
\begin{equation}\label{eqS}
 \rho(\mbox{H})\propto c\langle S\rangle^2,
\end{equation}
where $c$ is a constant and $\langle S \rangle \sim [n(\uparrow) - n(\downarrow)]$
is the field induced difference between the densities for electrons with spin parallel 
and anti-parallel to the field orientation if one assumes that the antiferromagnetism comes 
from a SDW state. On the other hand, $\langle S \rangle$ must be interpreted as the net
magnetization if one supposes that localized moments in the Fe atoms governs the magnetically 
ordered phase of the studied compounds. In any case, the field disturbs the cooperative spin alignment and should
increase the electron scattering rate. The fact that the MR increases as
a power law of the field with exponent  $b\sim$1.5 indicates that $\langle S \rangle$
increases sub-linearly as a function of H.%, with exponent around $3/4$.

% Another possibility to treat the MR is the existence of spin-polarized transport
% from the spin-mixing term of the Campbell-Fert two current model  \cite{fert1, fert2}:
% 
% \begin{equation}
%  \rho(T)=\frac{\rho_{\uparrow}\rho_{\downarrow}+\rho_{\uparrow\downarrow}(\rho_{\uparrow}+\rho_{\downarrow})}
%  {\rho_{\uparrow}+\rho_{\downarrow}+4\rho_{\uparrow\downarrow}},
% \end{equation}
% where $\rho_{\sigma}$ is the resistivity of the electrons with spin $\sigma$ and $\rho_{\uparrow\downarrow}$ is the  
% mixing-spin term. We argue that the applied magnetic field can produce some non-dissipative electron spin reversals favoring
% $\rho_{\uparrow\downarrow}$ and thus heightening the resistivity with increasing magnetic field.
% It was shown that at very low temperatures the spin-mixing term can be dominant \cite{fert1} doing the magnetoresistance
% pass  for a maximum. Looking at the graphs on Fig. \ref{MR_hall}, it seems that the MR vs. T curves for the doped
% samples tends to a maximum at the lower measured temperatures, near T = 20 K, thus reinforcing our hypothesis.
%  Since the MR of the pure sample ($x=0$)  does not
% show the same tendency than that of the doped ones,
% we  think that the spin-mixing factor is dominant in a lower temperature regime in that sample.
% This fact indicates that the pure sample has the  stiffest magnetic phase and
% that stiffness is softened by doping.

The occurrence of a magnetic contribution to the resistivity in our samples is also suggested
by results plotted in Fig. \ref{MR_T}.
In these plots, one observes that the amplitude of the MR as a function of $T$ is qualitatively reminiscent
of an order parameter that becomes non-zero in temperatures $T < T_{\text{SDW}}$. This behavior is independent 
of the applied field and is consistent with Eq. (\ref{eqS}). We are thus led to consider that the MR of
lightly substituted 122 Fe-pnictides is originated from dissipative electron scattering by spin excitations 
in the antiferromagnetic phase. This hypothesis contrasts with the previously proposed mechanisms based only on the two-band 
conduction model with large dominance of one type of carrier below $T_{\text{SDW}}$ \cite{albenque, fang, olariu}.

\begin{figure}[ht]
 \centering
  \includegraphics[keepaspectratio,width =8.5truecm]{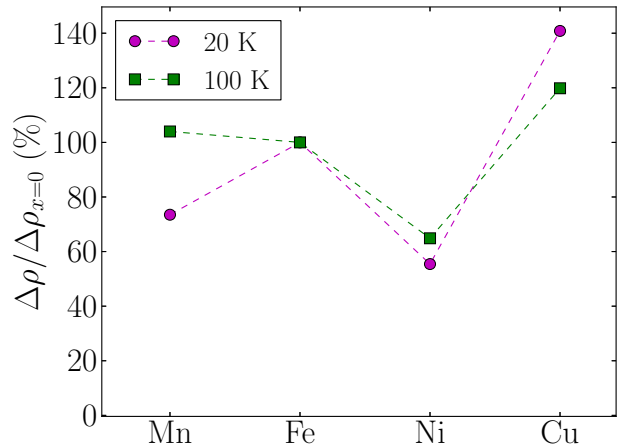}
  \caption{Absolute magnetoresistance  $\Delta\rho=\rho(\mbox{H}-\rho(0))$  for the measured samples in $T = 20$ K and 100 K
  under the applied field  $\mu_0H=9$ T.  The data are normalized to that for the pure sample.} \label{MR_compara}
\end{figure}

Figure \ref{MR_compara} shows $\Delta\rho=\rho(\text{H})-\rho(0)$ for our samples in temperatures 
$T = 20$ K and $T = 100$ K, and field \mbox{$\mu_0\text{H}=9$ T}. 
The data are normalized to that of the pure sample.
The results in \mbox{Fig. \ref{MR_compara}} roughly indicate that  the general
behavior of field-dependent electron scattering rate does not change drastically with 
chemical substitution and, at least at the concentration 
 range studied here, it does not show a systematic dependence on
 the substituting atom. 
% 
% When we compare the effect of different dopant atoms in reducing the MR we do not observe any systematic with
% the number of carriers added to the conducting planes. According to the inset of Fig. \ref{MR_compara}, Mn
% is the most effective dopant in reducing the MR at low temperatures and Co the less one.
% If we relate this with the fact that Co is the ion which substituted in the less proportion induces superconductivity
% in BaFe$_2$As$_2$, this reinforce the idea that the presence of magnetic fluctuations is important to develop superconductivity.
% However, the normalization in the inset of Fig. \ref{MR_compara} assumes that MR changes linearly with doping level
% in all the cases, but this has to be confirmed experimentally with complete series of samples
% for all dopant atoms and several levels of doping before giving any conclusion.

\begin{figure}
 \centering[ht]
  \includegraphics[keepaspectratio,width =8.5truecm]{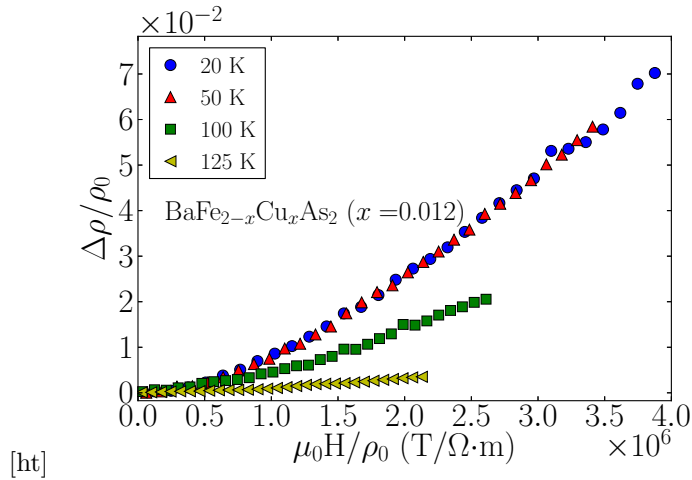}
  \caption{Representative plot showing the violation of the Kohler's rule in the magnetic phase 
  of the Cu-substituted sample for temperatures $T\gtrsim50$ K.} \label{kholer_Cu}
\end{figure}

Figure \ref{kholer_Cu} shows the Kohler plot for our Cu-substituted sample. The Kohler's rule is violated
in a large temperature interval inside the magnetically ordered phase (50 K $<T<T_\text{SDW}$). The Kohler's rule
is generally invalid in multiple band conductors. Moreover,
violations of Kohler's rule can be related to 
distinct scattering times of the charge carriers, with possible different temperature dependencies in the presence 
and absence of magnetic field and/or field dependent
scattering times \cite{mackenzie}. Thus, to our understanding, violations of Kohler's rule in our samples should be interpreted as
an indication of the presence of the carrier scattering by 
magnetic fluctuations in the ordered phase of our samples.
% On the other hand, disorder can be also considered to explaining the Kohler's rule violation.
% Disorder in our pure sample can explain the different results reported in Ref. \cite{gosh},
% where the Kohler rule was not verified in one sample of BaFe$_2$As$_2$
%  at any temperature (however it has to be taken into account that in Ref. \cite{gosh} the  magnetic field 
%  was applied parallel to the conducting planes while we just performed measurements with fields applied parallel to the $c-$axis).
% Disorder could also explain the fact of our 
% sample to have a $T_{\text{SDW}}\sim 135$ K, slightly lower than the commonly $T_{\text{SDW}}\sim 140$ K reported value  \cite{yi}.
% It is curious to note that as well as we verified the Kohler's rule at low temperatures in our samples,
% this rule was also recently verified in the pseudo-gap phase of Cuprate superconductors ~\cite{chan}.

\subsection{Hall effect}

\begin{figure}
 \centering
  \includegraphics[keepaspectratio,width =7truecm]{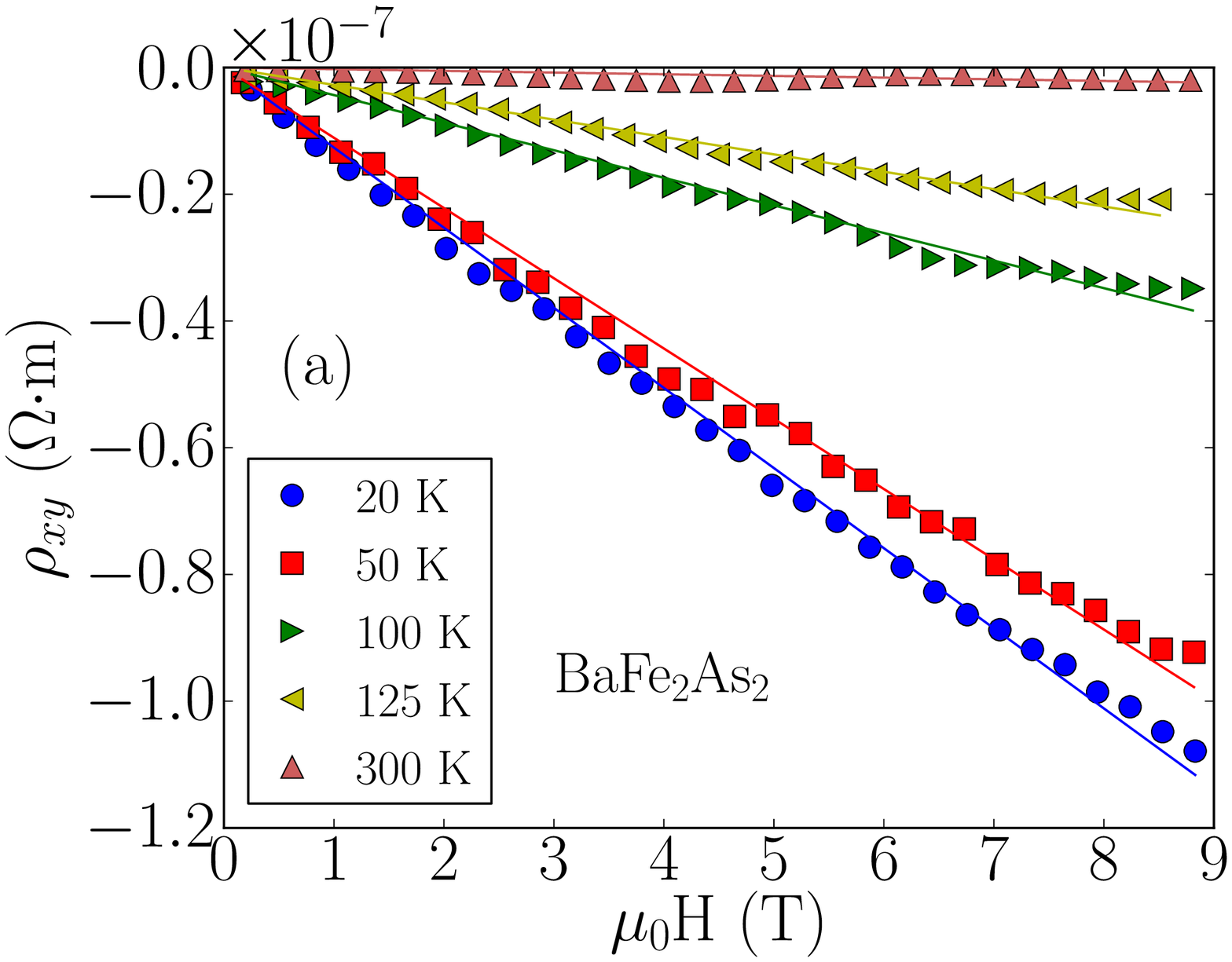}\\
\includegraphics[keepaspectratio,width =7truecm]{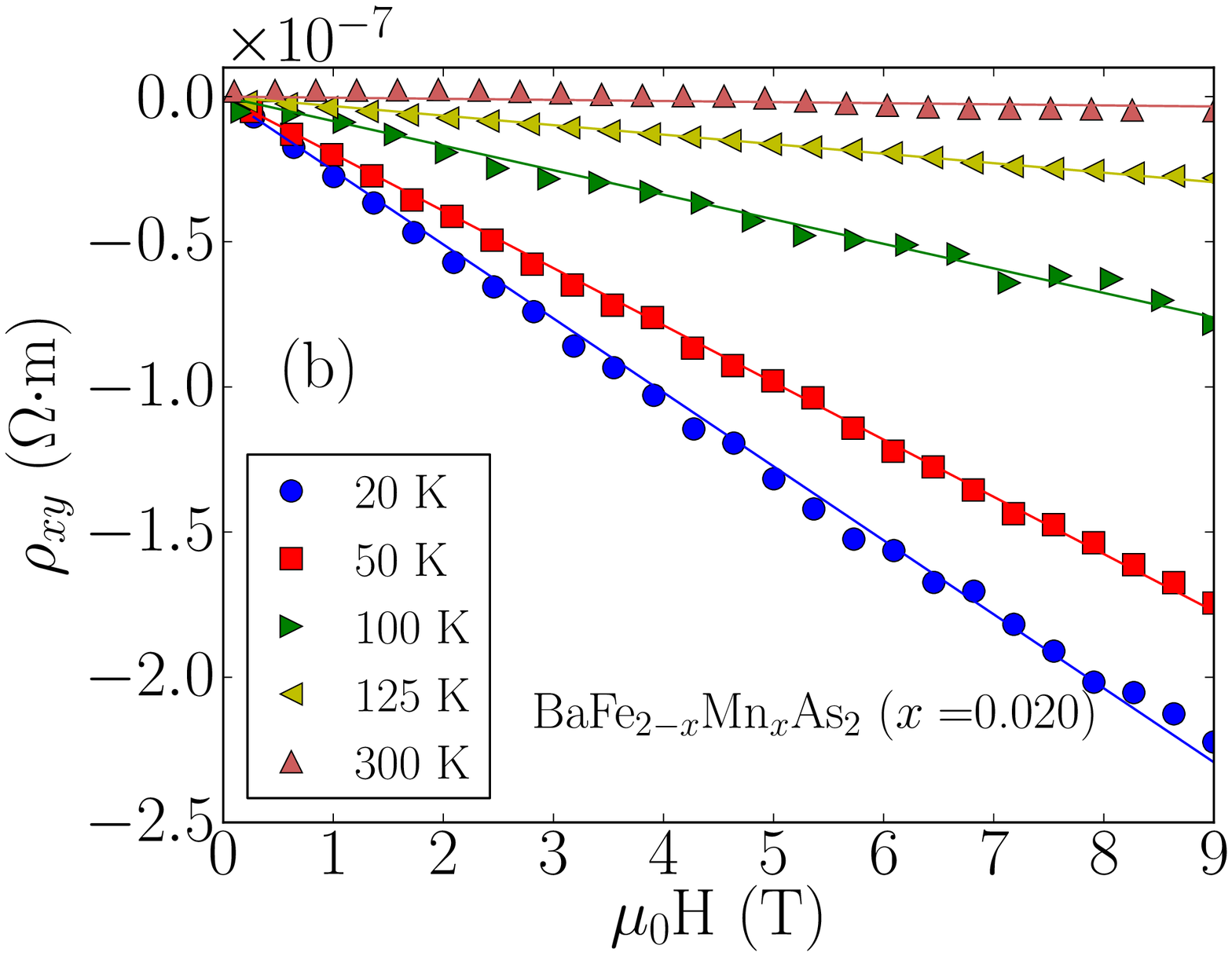}\\
  \includegraphics[keepaspectratio,width =7truecm]{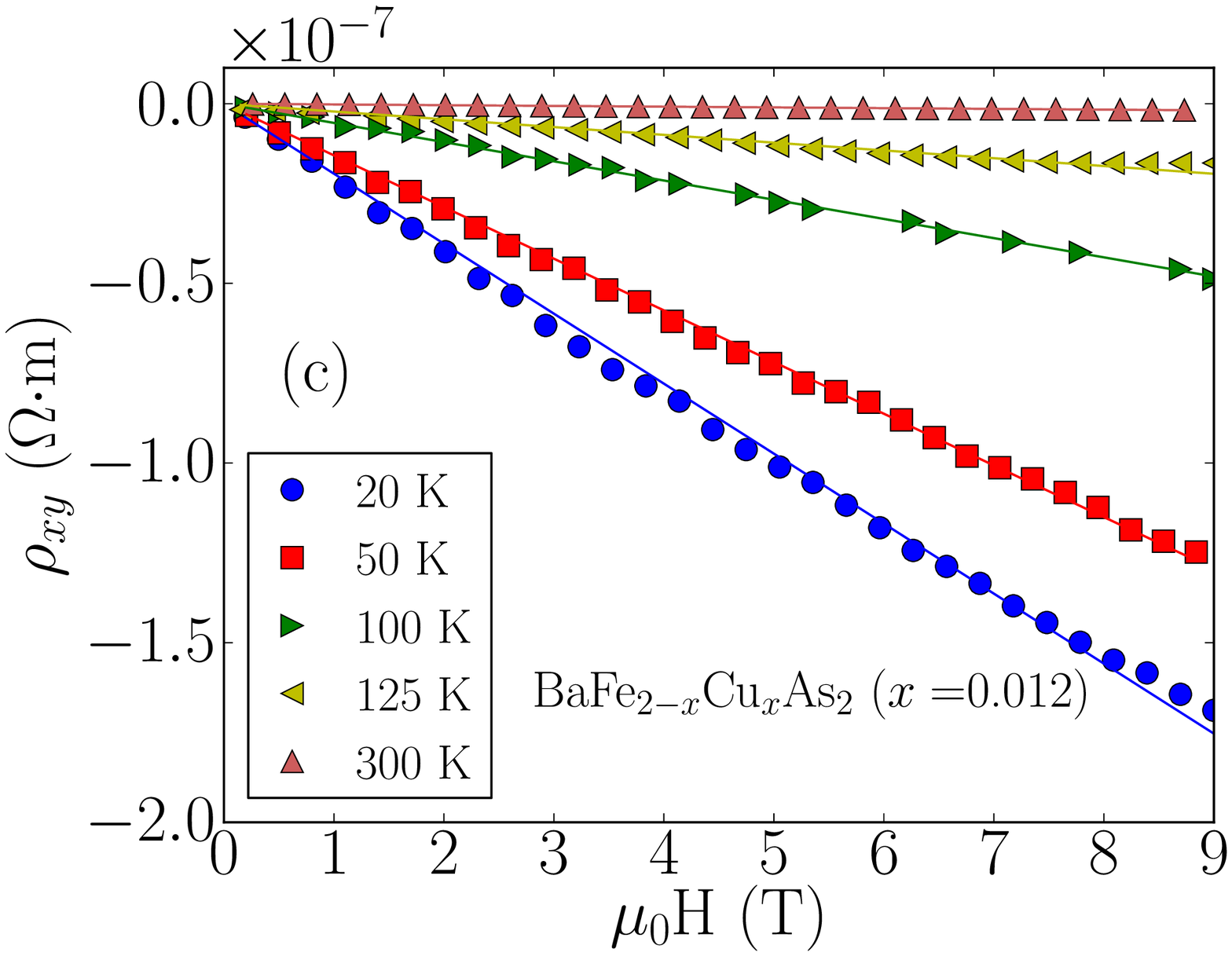}\\
\includegraphics[keepaspectratio,width =7truecm]{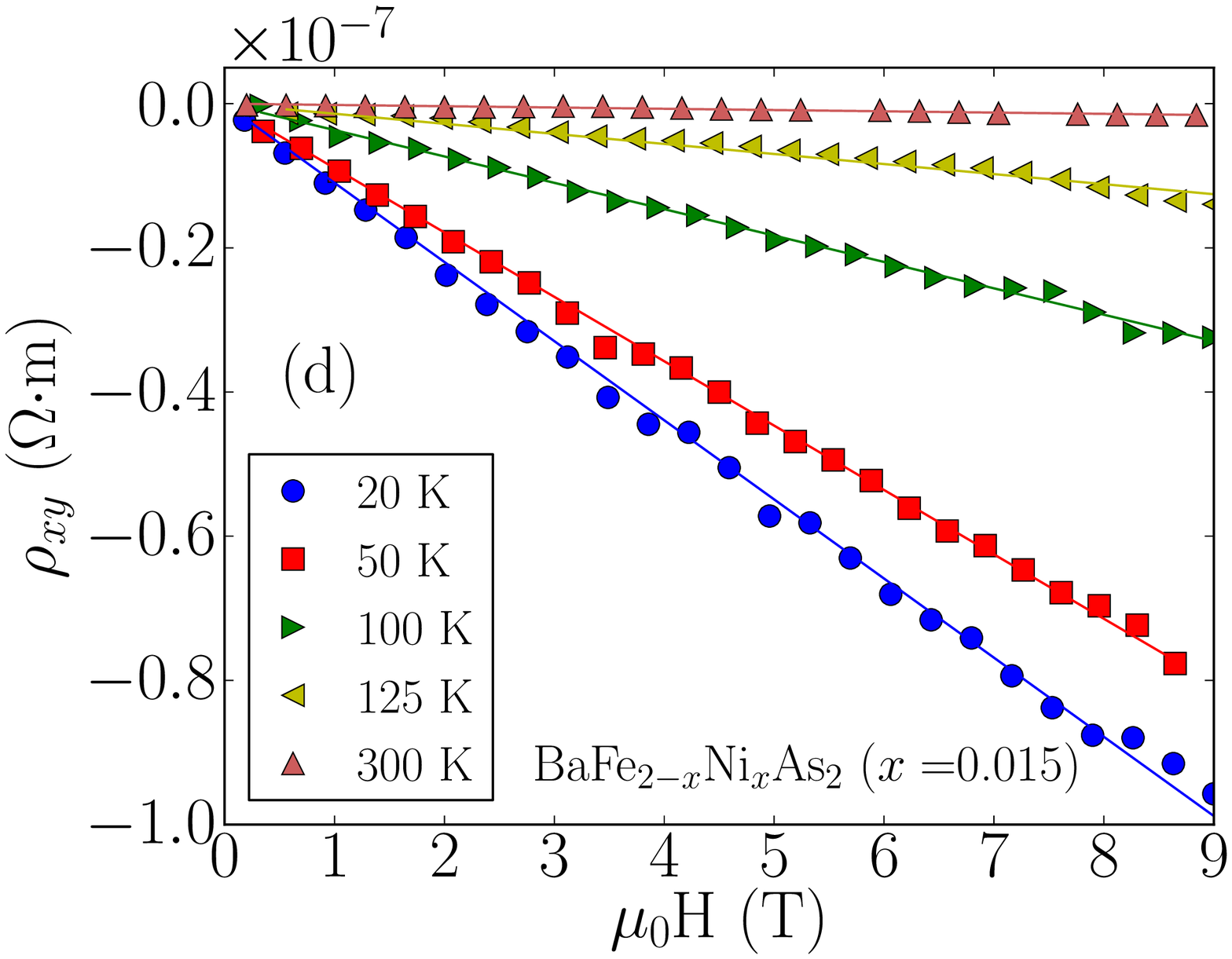}
  \caption{Hall resistivity  as a function of the applied magnetic field in several fixed temperatures. Solid lines are
  fits to straight lines.} \label{vs_field}
\end{figure}

\begin{figure}
 \centering
     \includegraphics[keepaspectratio,width =7truecm]{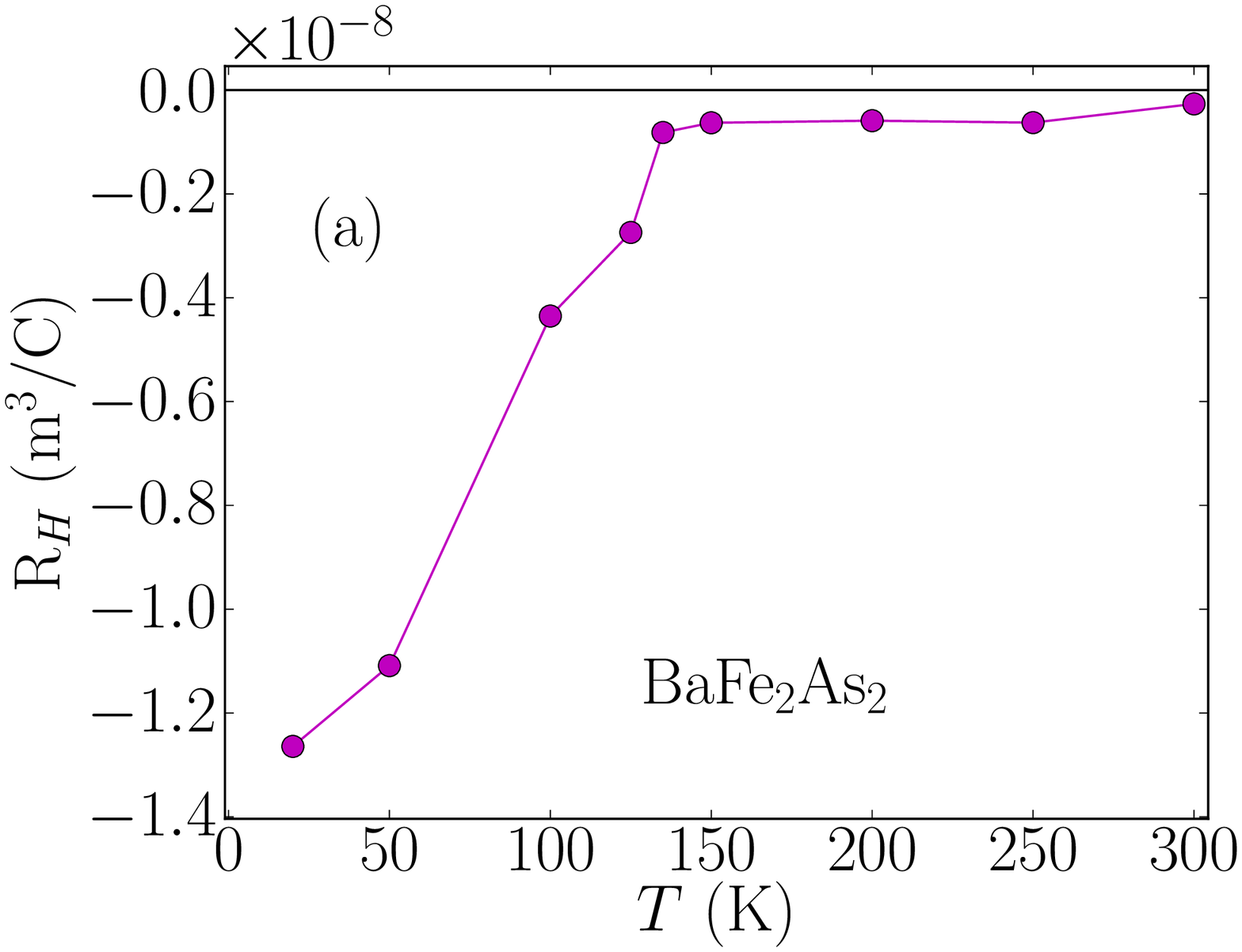}\\
   \includegraphics[keepaspectratio,width =7truecm]{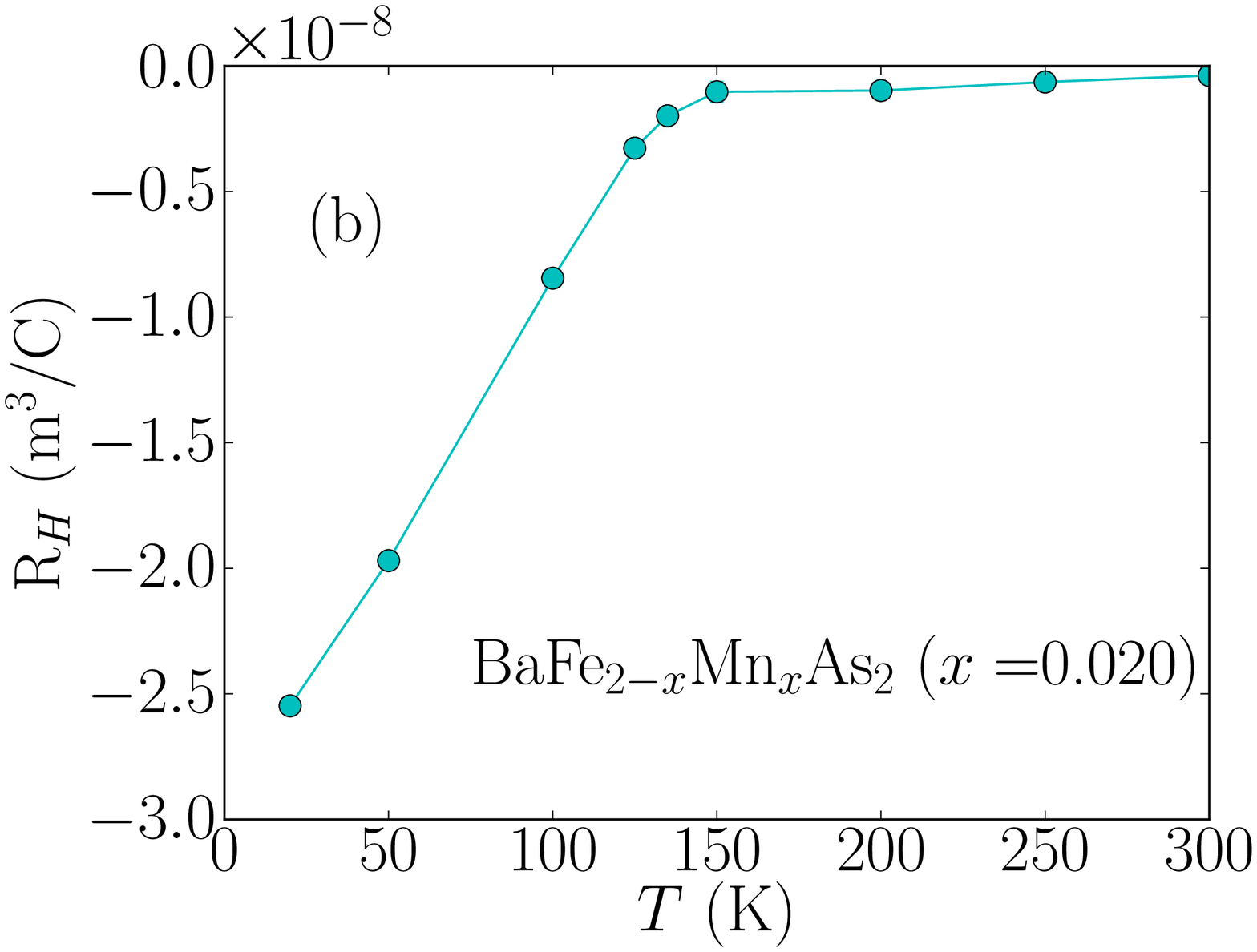} \\
   \includegraphics[keepaspectratio,width =7truecm]{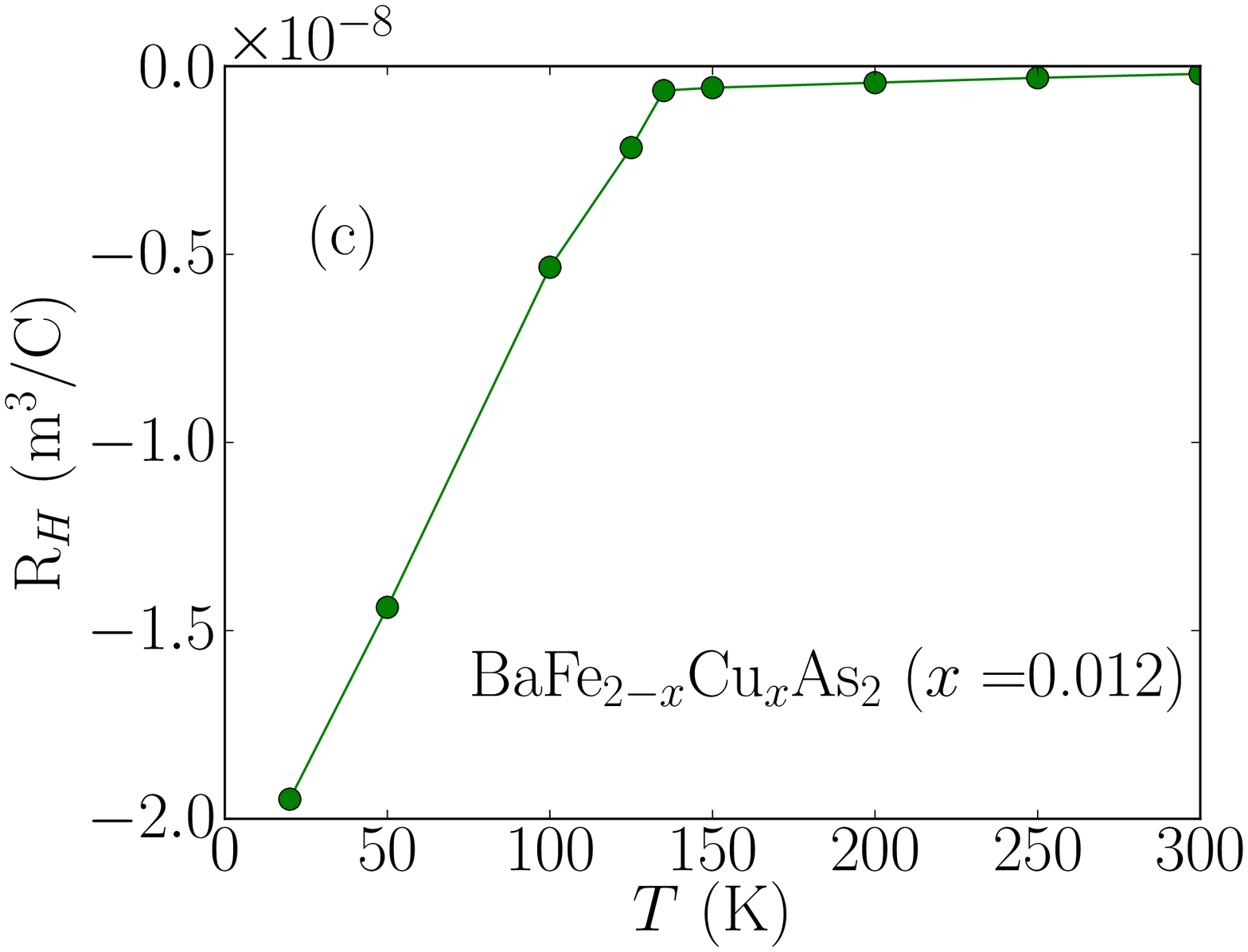} \\
  \includegraphics[keepaspectratio,width =7truecm]{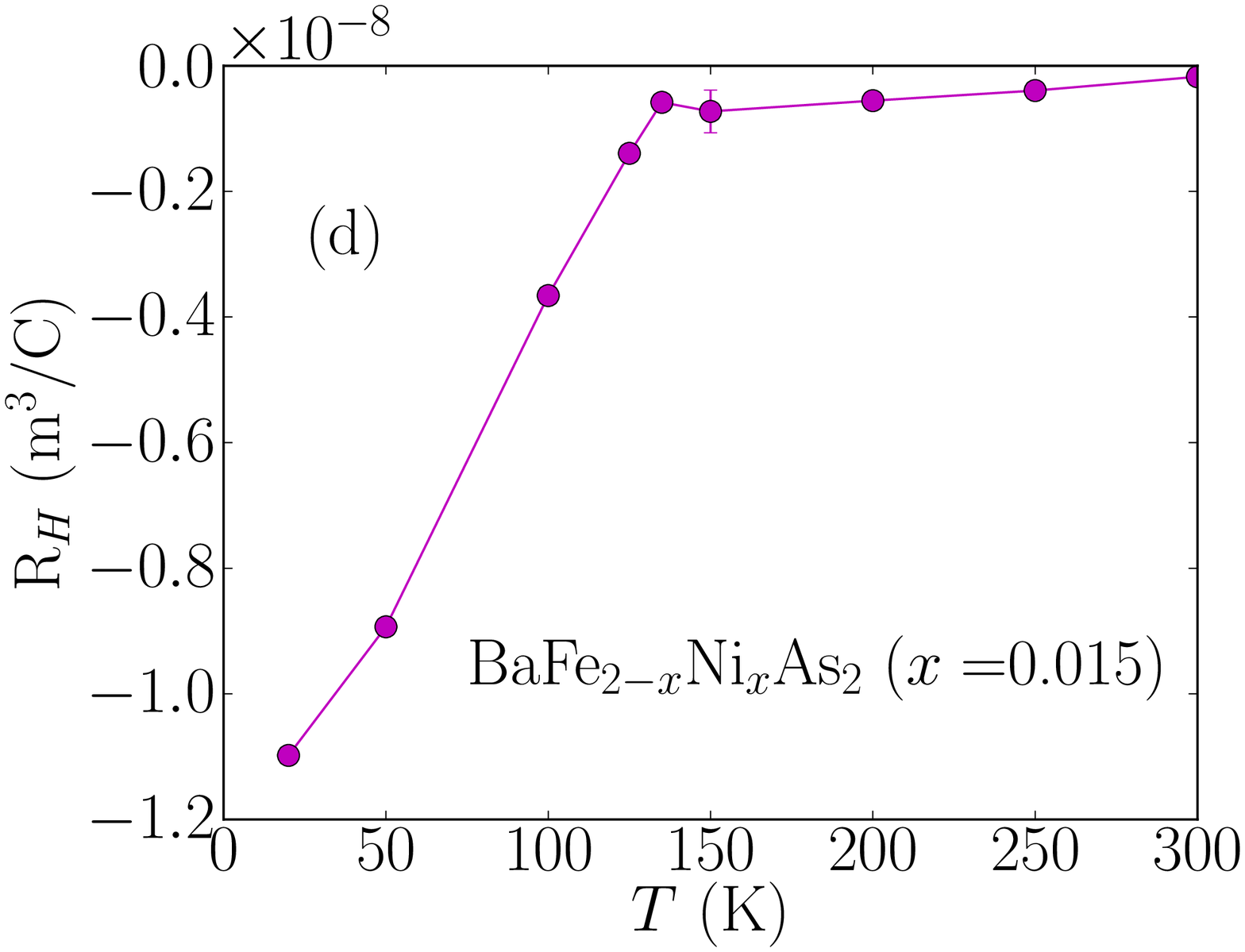}
  \caption{Hall coefficient as a function of the temperature. The points represent the $R_H$ value for each temperature was obtained
  from the slope of the straight line fits in Fig. \ref{vs_field}.} \label{vs_temperature}
\end{figure}

Figures \ref{vs_field} and \ref{vs_temperature} show the results from the Hall effect experiments carried out in the studied 
samples. In \mbox{Fig. \ref{vs_field}} the Hall resistivity $\rho_{xy}$ is presented as a function of the magnetic field at
several fixed temperatures. In all cases $\rho_{xy}$ is a linear function of $\mu_0\text{H}$. 
In this respect, our pure sample differs from others reported in literature \cite{kuo} where the  $\rho_{xy}$ vs. $\mu_0\text{H}$
curves present a slight positive curvature for fields $\mu_0\text{H}\gtrsim2$ T. We speculate that our pure specimen
presents some kind of disorder which has the same effect in $\rho_{xy}$ as the addition of impurities.
This could also explain the fact that our pure
sample to have a $T_{\text{SDW}}\sim 135$ K, slightly lower than the commonly reported value $T_{\text{SDW}}\sim 140$ K.
In \mbox{Fig. \ref{vs_temperature}} the Hall coefficient $R_H$ is shown as a  function of the temperature
for all samples. 
The Hall coefficient was obtained from the slope of the straight lines fitted to the $\rho_{xy}$ vs. $\mu_0\text{H}$ data in
\mbox{Fig. \ref{vs_field}}.

In Fig. \ref{vs_temperature} one observes that  $R_H$ is negative, small and
weakly temperature dependent in the paramagnetic
region \mbox{($T>T_{\text{SDW}}$)}. At the magnetic ordering temperature,
a remarkable change occurs in the Hall coefficient: while
remaining negative, $R_H$   becomes strongly temperature dependent, 
so that its magnitude increases roughly linearly by more than one order of magnitude as $T$ decreases toward zero.
The negative sign
of $R_H$ in the whole studied  temperature range 
indicates that, at these doping levels, the transport is  dominated by electrons in all the studied samples.
The fact that $|R_H(T)|$ decreases almost linearly to small 
values when $T$ approaches $T_{\text{SDW}}$ from below, and shows a remarkable change of behavior at this temperature,
is in agreement with the temperature dependence of the magnetoresistance amplitude
shown in Fig. \ref{MR_T}. Also,  the weak dependence of the $R_H$ magnitude on the imputity atom is
remarkable in the results of \mbox{Fig. \ref{vs_temperature}}. Consequently, in accordance with the resistivity and magnetoresistance
results, the Hall effect data in \mbox{Fig. \ref{vs_temperature}} suggest that substituting atoms at the Fe 
site of the 122 Fe-pnictides act mostly as scattering centers than as effective dopants, 
at least within the concentration limit studied here. 
This result is in agreement with previous reports which suggested that the main role of the distinct chemical 
substitutions in the Ba122 family is to provide local distortions near the Fe site \cite{rosa, garitezi2, rosa2},
so that subtle variations of the
structural parameters play a major role to explain the phyical properties of the Fe-pnictides \cite{yin2}.

The behavior of $R_H$ in the paramagnetic region is in accordance
with the expectations for conduction by two currents, consistently with the previously described magnetoresistance results.
According to the two-band conduction model the Hall coefficient is written as \cite{ziman}:
\begin{equation}\label{eqRH_2b}
 R_H\approx\frac{\sigma_h\mu_h-\sigma_e\mu_e}{(\sigma_h+\sigma_e)^2}.
\end{equation}
The small values of $R_H$ in the paramagnetic regime indicate nearly
compensated transport by holes and electrons; it means that not only mobilities, but the electron and hole conductivities 
are close to each other  in the paramagnetic phase in all cases.

In the paramagnetic region, where the effect of  magnetic order can be neglected,
authors in Ref. \cite{fanfarillo} theoretically proposed 
that the variation of $R_H$ with the temperature can be
understood by introducing an additional term to the Hall effect which origin is beyond the Boltzmann approximation.
This supplementary term corresponds to
vertex corrections to  the quasi-particle currents. These corrections are argued to be necessary because of the mixing  
produced by the exchange of spin fluctuations between the 
electron ($e$) and hole ($h$) bands.
Others suggest that the  temperature dependence of $R_H$ in $T>T_{\text{SDW}}$ is associated with
the momentum-dependent scattering off spin fluctuations and the ellipticity of the electron pockets  \cite{breitkreiz1, breitkreiz2}.
% This give rise to anisotropies in the interband scatterings generating instabilities and  transport anomalies
% as the opposite drift direction of the minority carriers .

On the other hand, explanations for the temperature  dependence of $R_H$ in the magnetically ordered region
include, in most cases, a severe
Fermi surface reconstruction  generated by the orthorhombic distortion accompanied by the SDW ordering.
While some authors consider
the reduction of the charge carriers density \cite{albenque, yin2, yin}, 
others propose that a reduction of the hole mobility or an enhancement of electron mobility is the 
dominant phenomenon \cite{fang, olariu}.

% which drastically modify the charge carriers density. 
%However the same arguments can not be taken
% as the only contribution  to explain the magneto-transport phenomena because 
% its applicability depends strictly on the value of the SDW potential 
% which we think may be largely modified by the magnetic field. For example, results in Fig. \ref{vs_temperature}
% show that $R_H$ continues to drop in temperatures far below T$_{SDW}$ in all cases and do not stabilize 
% as would be expected from one sharp Lifshitz transition. 
% On the other hand, a substantial augmentation of the mobility might be envisaged if a band of Dirac-type 
% carriers stabilizes at the Fermi level below T$_{SDW}$, as suggested to occur in the undoped compound by ARPES experiments 
% and theoretical calculations \cite{richard, ran, harrison}. 
% However, these explanations and the former ones
% encounter difficulties to explain the strong temperature dependence of the Hall coefficient 
% in the magnetically ordered state of the doped samples.
% The argument of 
% reduction of charge carriers is criticized because it does not take into account the holes contribution to
% the transport in any temperature regime. Conversely, the mobility approach
% requires the ``freezing'' or localization of the hole carriers promoted by spin fluctuations \cite{fang}; 
% the criticism of this
% argument is that  such  a mechanism should affect the mobilities of both electron and holes,
% and it is not clear what does that only the holes be frozen at low temperatures.

Here, we propose that  contributions related to magnetic excitations are also important 
to explain  the Hall effect in the ordered state of
 122 Fe-pnictides.
In particular, we associate an extraordinary, or anomalous contribution to the Hall effect  in addition
to the ordinary one described by Eq. \ref{eqRH_2b}. Since the ordinary contribution is very small due to 
the almost compensated two-band conduction, we suggest that  the $R_H$ magnitude in the temperature region
below $T_{\text{SDW}}$ largely comes from the anomalous term.

%\subsection{Anomalous Hall Effect}
 
The anomalous contribution to the Hall effect (AHE) is
mostly associated with ferromagnetic metals. It is parametrized
as a linear function of the magnetization, so that the total Hall resistivity is  written as \cite{sinitsyn}:
\begin{equation}\label{eq_anomalo}
 \rho_{xy}(\text{H})= \mu_0(R_0\text{H}+R_S\text{M}),
\end{equation}
where $\mu_0$ is the vacuum permeability,
$R_0$ is the ordinary Hall coefficient,  M is the magnetization and $R_S$ is the anomalous
Hall coefficient. The first term in the right side of Eq. (\ref{eq_anomalo})
represents the ordinary Hall effect due to the Lorentz force and the second one is the anomalous contribution.
At first glance, since antiferromagnetic metals 
have zero net spontaneous magnetization, Eq. (\ref{eq_anomalo}) rules out the
description of the Hall effect in  these materials. 
Consequently, up to now an anomalous contribution to the Hall effect has not been taken into 
account for explaining the behavior of the magneto-transport phenomena in 
Fe-pnictides. 
However, we must note that antiferromagnetic metals can also develop an AHE induced by  non-collinear spin 
structures \cite{kubler, chen}, %, magnetic frustration and spin canting. 
and even non-frustrated collinear lattices may develop anomalous contributions to the Hall
current \cite{chendong} % .  That contributions come from lattice distortions  which
%change the hopping amplitude of the itinerant electrons \cite{chendong}. 
as observed in the U$_2$PdGa$_3$ \cite{tran} and Nd$_{1-x}$Ca$_x$B$_6$ \cite{jolanta} compounds.

\mbox{Equation (\ref{eq_anomalo})} is an empirical relation which should not be taken as universal, 
neither suitable for all materials \cite{sinitsyn}. However, since
antiferromagnetic materials can develop a field-induced magnetization, which  may be approximately written as
M $ = \chi_{eff}(T)$H, where $\chi_{eff}(T)$ is a  temperature dependent effective susceptibility, here  we  assume 
the validity of Eq. (\ref{eq_anomalo}) to describe the results in Figs. \ref{vs_field} and \ref{vs_temperature}.
Then we write:
\begin{equation}\label{eq_anomalo2}
 \rho_{xy}(\text{H})= \mu_0R_H\text{H},
\end{equation}
where
\begin{equation}\label{eq_RH_anomalo}
 R_H = R_0 + \chi_{eff} (T) R_S (T).       
\end{equation}
% The results in Figs. \ref{vs_field} and  \ref{vs_temperature}
% are fully consistent with the description given by Eqs. (\ref{eq_anomalo2}) and (\ref{eq_RH_anomalo}).
     
Based on the results shown in  Fig. \ref{vs_temperature}, we suppose that $R_H \simeq R_0$
in the paramagnetic phase. A small contribution from an anomalous term coming from scattering
by magnetic impurities, and 
corrections as those considered in Refs. \cite{fanfarillo}, \cite{breitkreiz1} and \cite{breitkreiz2},  might
lead to the temperature dependence of $R_H$ in the $T>T_{\text{SDW}}$ region. 
As a working hypothesis, we assume that $R_0$ remains small ($R_0\sim10^{-9}$ C/m$^3$ in all cases) below $T_{\text{SDW}}$,
so that the strong temperature dependence of $R_H$ in this temperature range is due to
the anomalous Hall contribution.
Thus, we suppose that $R_H(T) \simeq \chi_{eff}(T)R_S(T)$  for $T<T_{\text{SDW}}$.

Key ingredients to develop AHE are the multi-orbital character of the charge carriers, 
spin-orbit interaction and time-reversal symmetry breaking. The multi-orbital character of the carriers in
the Fe-pnictides is already known. The existence of spin-orbit interaction in these compounds was theoretically  predicted 
\cite{fernadesv}, and experimentally corroborated  by ARPES measurements \cite{borisenko}.
The application of a magnetic field naturally breaks the time-reversal-symmetry, but 
an intrinsically broken time-reversal symmetry related to the particular magnetic ordering in the 122
compounds should be present in order to generate an enhanced AHE.
At this point, one might consider that some ``hidden'' magnetic order   \cite{rodrigues} 
accompanying the SDW state plays 
a role to explain the broken time-reversal symmetry in the 122 
Fe-pnictides. %The idea of a hidden magnetic order coexisting with the SDW phase 
%was introduced in Ref. \cite{rodrigues}. The authors 
%argue that the hidden order could be produced by Heisenberg spin exchanges between different 3$d$ 
%orbitals in neighboring iron atoms. This hidden order can be a N\'eel or
%a ferromagnetic order if the exchange interaction is sufficiently frustrated.}.

Taking the above considerations into account, we  propose
that  the inclusion of an anomalous term for explaining the
Hall effect results in the magentic state of our samples is a  reasonable assumption.
Of course this statement leads to the issue of determining the origin of the AHE in these materials.
Four mechanisms are conventionally assumed to produce AHE in ferromagnetic 
materials \cite{sinitsyn, nagaosa}.
The so-called intrinsic mechanisms are related to Berry phase 
effects on the Bloch wave-functions of the charge carriers under the influence of spin-orbit scattering.
Two different possibilities may occur: the Berry phase may accumulate in the reciprocal space 
giving origin to a dissipationless transverse current \cite{karpus}, or the carriers can accumulate a Berry 
phase in the real space due to canting of localized spins. The former  is called the Karplus-Luttinger \mbox{(K-L)}
contribution  \cite{karpus} and the last one
is known as the spin-chirality mechanism \cite{taguchi}.
Two extrinsic mechanisms for AHE are due to the interaction of the charge carriers with atomic magnetic scattering centers
in the material and depend on the concentration of those single-ion moments; these are the skew-scattering
and side-jump contributions \cite{nagaosa}. 

Identifying experimentally the contribution 
of each mechanism to the AHE of a given material may be difficult. In general, a useful procedure is 
to plot the anomalous Hall coefficient as a function of the longitudinal resistivity assuming that 
these quantities relate to each other through a simple power-law  with the form $R_S \propto \rho_{xx}^\beta$.
The K-L theory predicts that $\beta=$ 2 \cite{nagaosa, karpus}. For the skew-scattering mechanism the exponent is 
$\beta=$ 1  provided that $\rho_{xx}$ increases linearly with the concentration of single-ion moments.
For the side-jump one also expects $\beta=$ 2, since the effect should depend on the square of
the single-ion moment concentration \cite{nagaosa}. As a general trend, the  K-L mechanism 
is expected to describe the AHE in systems where the magnetic moments distribute periodically in the lattice. 
Skew-scattering is dominant in dilute magnetic alloys while side-jump should prevail in concentrated magnetic alloys.
In the case of our experiments, since  $\chi_{eff}(T)$ data are not available for our samples, we are unable to 
single out the value of $R_S(T)$ from the $R_H$ value.  However, measurements of M/H existing in the litterature for pure and
substituted 122
compounds, with similar and higher content of the Mn impurity ~\cite{thaler}
show a temperature dependence that is reminicent of the $R_H$ results in Fig. \ref{vs_temperature}(b). In other words, 
the temperature dependence of $R_H$ can be largely dictated by $\chi_{eff}(T)$.

Even so, if we consider that $\chi_{eff}(T)$ is approximattelly constant in a large temperature interval
inside the magnetic region in pure and very low substituted samples ~\cite{sefat2, thaler},
we obtain the relation $R_H \propto R_S(T) \propto \rho_{xx}^\beta$. Thus, in this case we can evaluate
the dependence of  $R_H$, or equivalently of $\rho_{xy}$, on $\rho_{xx}$.
The results for $\rho_{xx}$  in Fig. \ref{RT} and $\rho_{xy}$  in Fig. \ref{vs_field}
show that the longitudinal and Hall resistivities in the samples  studied here are in anti-correlation 
in the magnetically ordered phase; 
that is, $\rho_{xx}$ increases while $\rho_{xy}$  decreases 
when plotted as functions of the temperature in the range 
$T<T_{\text{SDW}}$. Then, one can not relate the Hall resistivity to a single power-law of the longitudinal resistivity
with exponent $\beta=$ 1 or $\beta=$ 2.  
This fact suggest that the AHE
in our samples could be due to the spin-chirality mechanism \cite{kawamura}. 
The theory for this effect does not predict any correlation of $R_S$ with the longitudinal
resistivity \cite{tatara}. When identified experimentally, the contribution of the spin-chirality to the AHE does not 
show  any obvious dependence with $\rho_{xx}(T)$ \cite{wolff}.
The spin-chirality mechanism implies that canting of local spins with respect to the magnetization must occur, so that
the triple product ($\overrightarrow{S_i}\bullet\overrightarrow{S_j}\times\overrightarrow{S_k}$) of 
neighboring spins is non-zero. This canting may be static and related to local disorder. 
Because of the presence of microscopical defects as twinning, dislocations and vacancies, 
lattice distortions are likely to occur in profusion even in single crystal samples 
of the 122 Fe-pnictides (especially in the presence of chemical substitution)  \cite{tran}; consequently, spin-chiralities 
can be good possible generators of the AHE in this systems. However, 
the mechanism of spin-chiralities can also be relevant in collinear spin systems, provided that  the inelastic scattering rate
for conduction electrons is larger than the relaxation rate for the spin excitations \cite{wolff}.

\begin{figure}
 \includegraphics[keepaspectratio,width =7.2truecm]{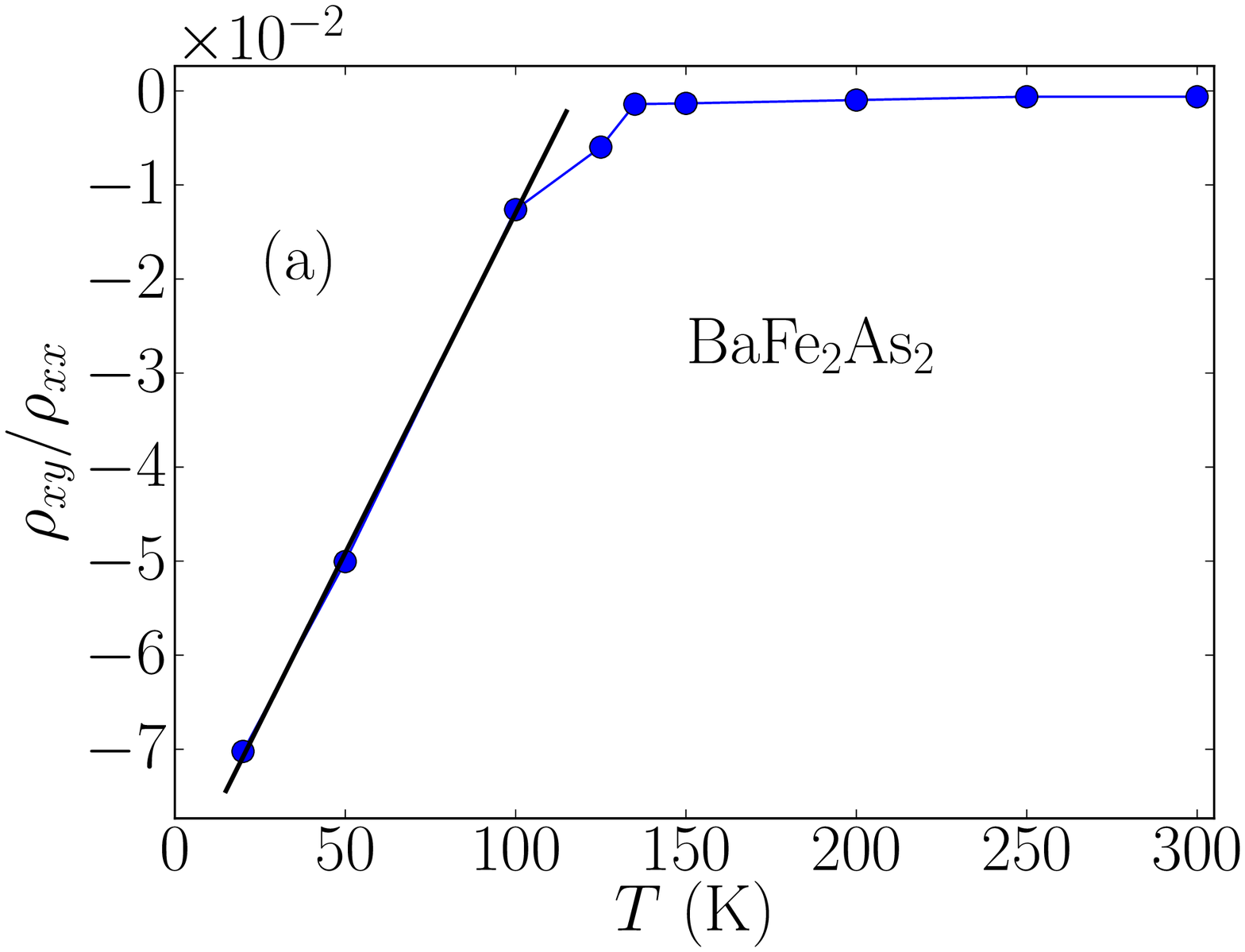}
 \includegraphics[keepaspectratio,width =7.2truecm]{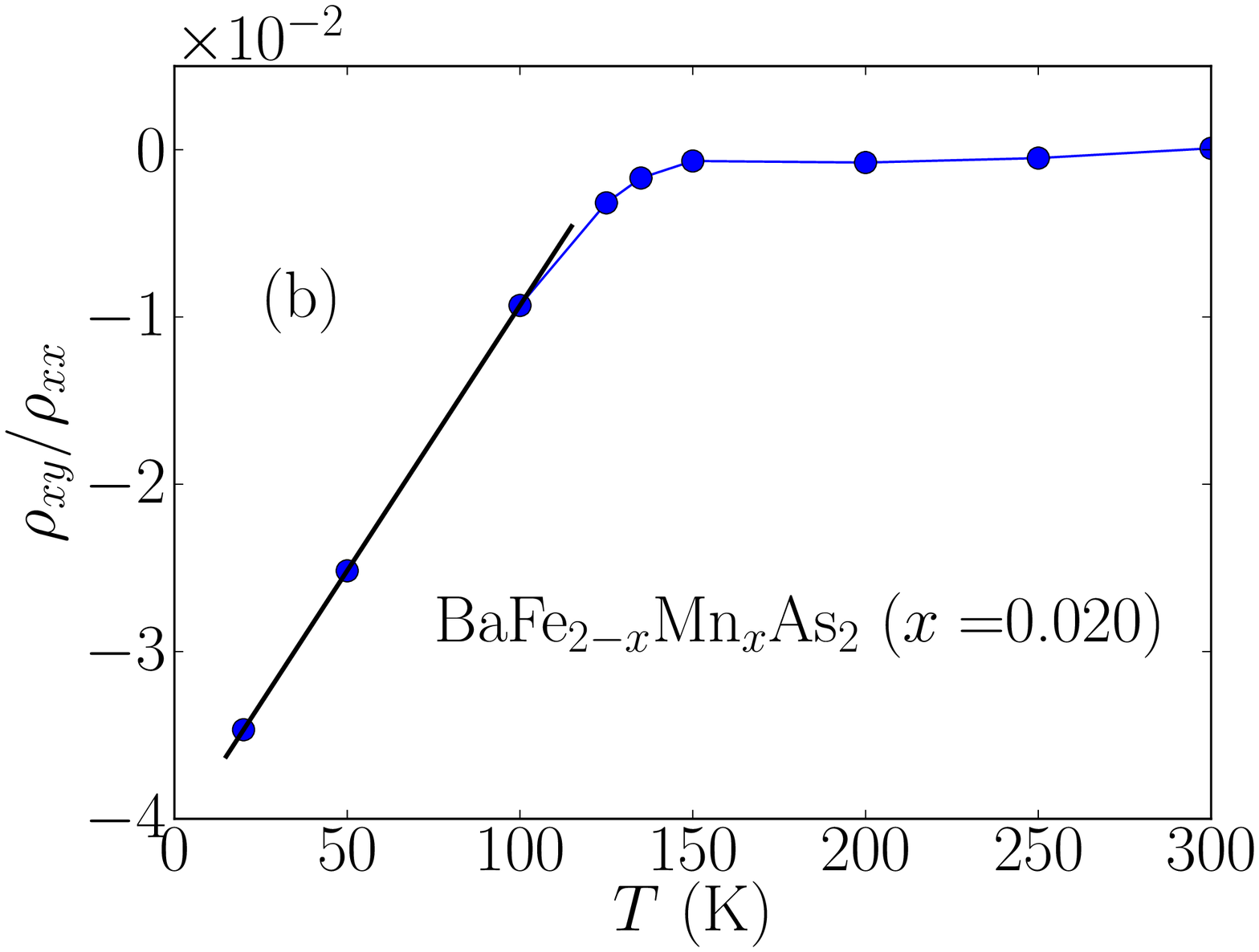}\\
 \includegraphics[keepaspectratio,width =7.4truecm]{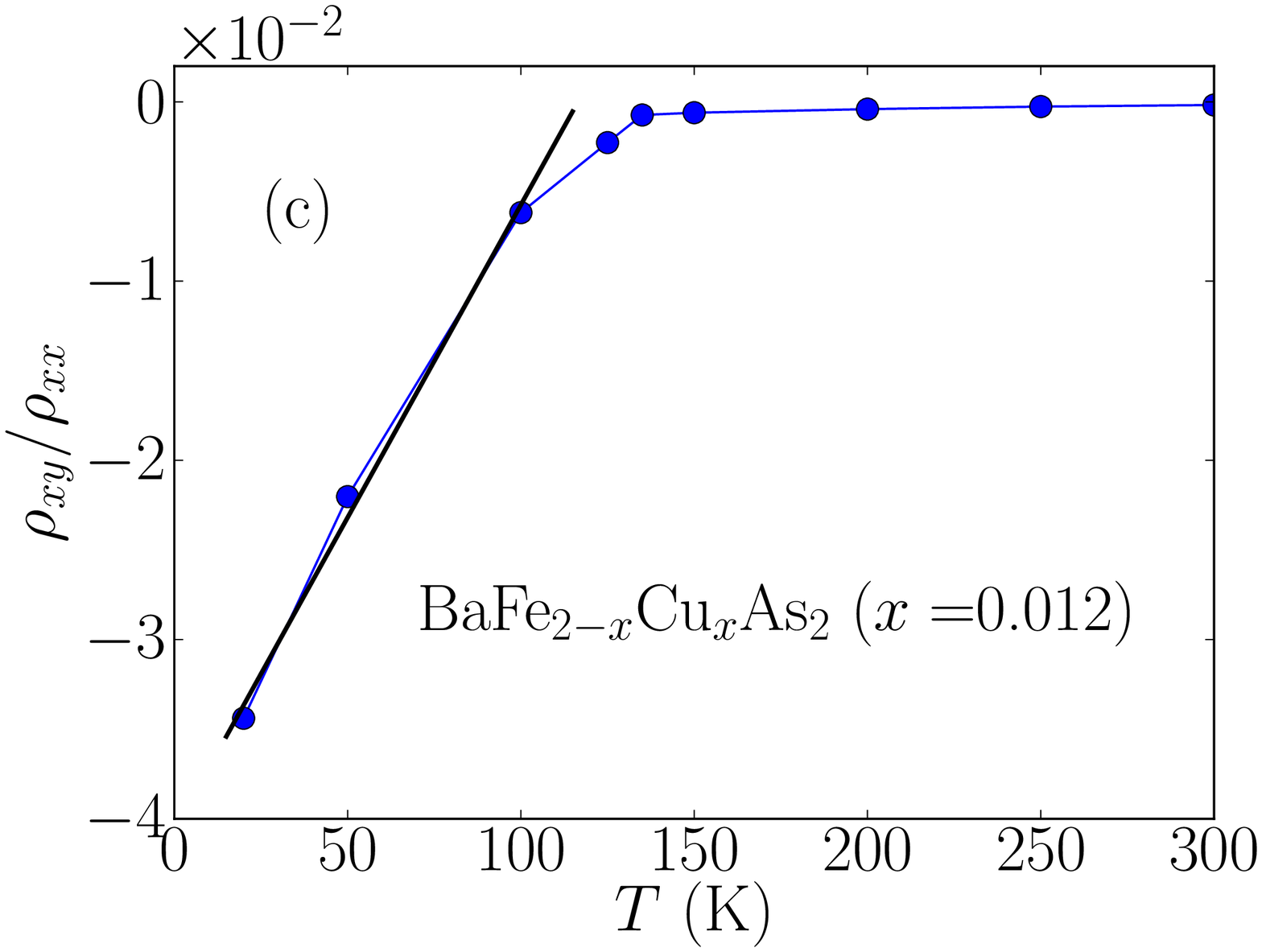}
 \includegraphics[keepaspectratio,width =7.4truecm]{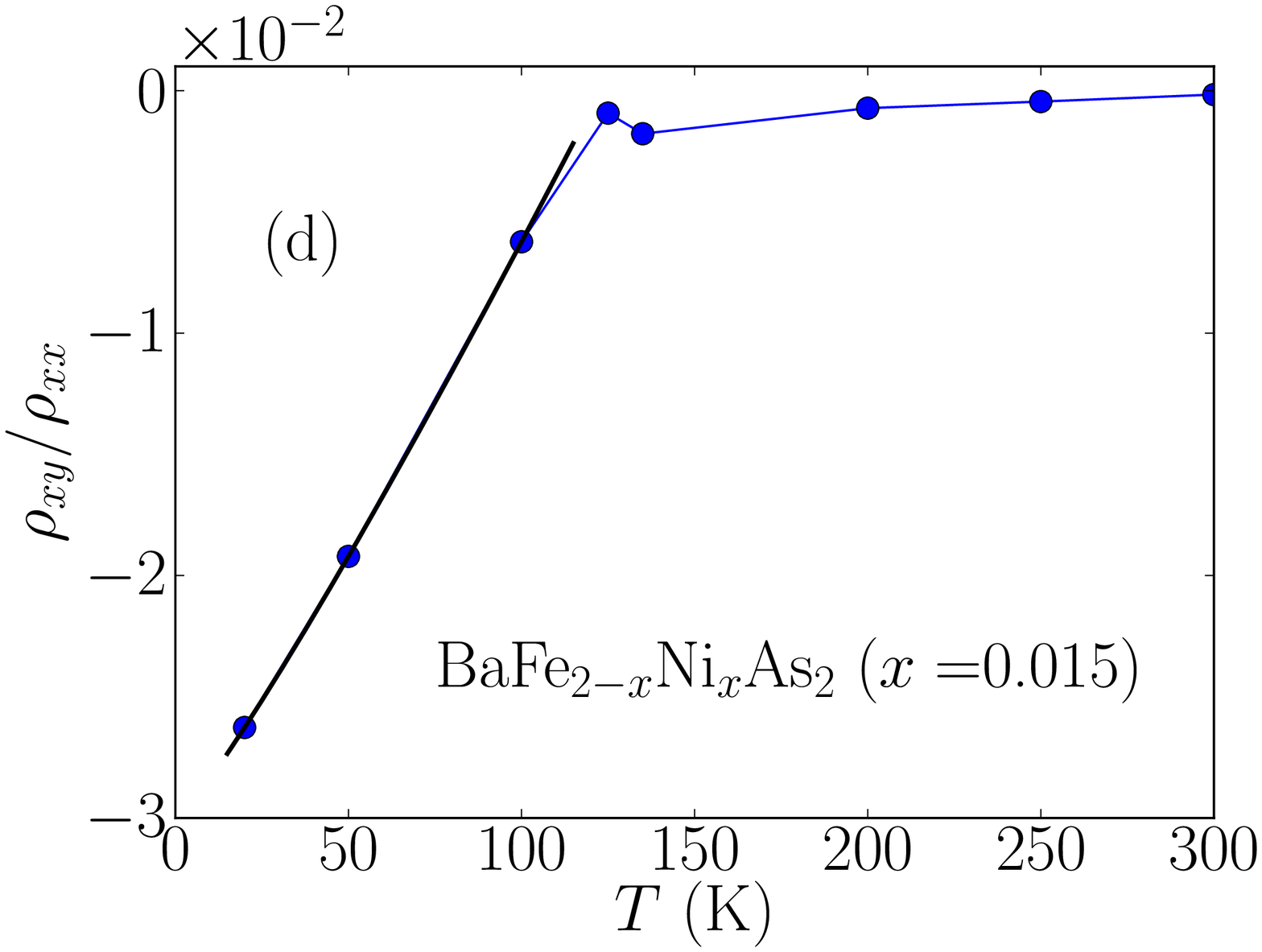}
 \caption{Tangent of the Hall angle for an applied field \mbox{$\mu_0\text{H}=4$ T} as a function of the temperature for the studied samples.
 Solid straight lines in the low temperature region are fits of the experimental points to Eq. (\ref{eq_tanT}).} \label{rxy_rxx}
\end{figure}

In order to obtain an additional insight on the Hall effect in the low temperature region, 
in Fig. \ref{rxy_rxx} we plot the tangent of the Hall angle ($\tan\Theta_H = \rho_{xy} / \rho_{xx}$)
as a function of the temperature for  the studied samples. This parametrization for the Hall
 effect is of limited value here since the residual resistivity represents a
 large contribution to the total resistivity of these samples in the whole temperature range. However, a remarkably simple result 
is obtained in the region $T<T_{\text{SDW}}$, where $\tan\Theta_H$ behaves as a linear function of the temperature as:
\begin{equation}\label{eq_tanT}
  \tan\Theta_H=\alpha-\beta T.
\end{equation}

Table \ref{Tan_parameters} lists the parameters $\alpha$ and $\beta$ obtained from the best fits
to Eq. (\ref{eq_tanT}) in the low temperature range in  Fig. \ref{rxy_rxx}.  

\begin{table}[h]
\centering
\caption{Fitting parameters for the tangent of the Hall angle in the magnetic phase of the 
listed systems according to Eq. (\ref{eq_tanT}).}\label{Tan_parameters}
\begin{tabular}{|c|c|c|}\hline 
\textbf{Sample} & $-\alpha (10^{-2})$ &  $\beta (10^{-4})$ (K$^{-1}$) \\%$T_{\text{SDW}}\pm1$ (K) &
\hline 
$x=0$ &  8.5 & 7.2\\ 
\hline 
Mn$_{0.020\pm0.006}$ & 4.1 & 3.2\\
\hline 
Cu$_{0.012\pm0.004}$ & 4.1 & 3.5\\
\hline 
Ni$_{0.015\pm0.005}$ & 3.2  & 2.5\\
\hline 
\end{tabular}
\end{table}

The behavior of the tangent of the Hall angle  in the magnetically ordered region as described by Eq. (\ref{eq_tanT}),
can not be explained with the single conduction band model. In this case, one expects that 
$\tan\Theta_H=\omega_c\tau$, where  $\omega_c$ is the cyclotron frequency and  $\tau$
is the electron transport relaxation time. In the simplest two-band model, one may write 
$\tan\Theta_H=(\sigma_h\omega_h\tau_h-\sigma_e\omega_e\tau_e)/(\sigma_h+\sigma_e)$, where $\omega_{h(e)}$ 
and $\tau_{h(e)}$ are the cyclotron frequencies and relaxation times for holes (electrons), respectively.
Unless assuming a rather unusual and somewhat conflicting temperature dependencies for the relaxation 
times and/or conductivities for holes and electrons, it is impossible to describe results for
$\tan\Theta_H$ in the region $T<T_{\text{SDW}}$ with basis solely on two-band conduction. 
Then, results in Fig. \ref{rxy_rxx} also  led us
to suppose that the Hall effect in the magnetically ordered state of the undoped and slightly doped 122 Fe-pnictides 
is mostly due to an anomalous component. This component behaves differently 
than those identified in ferromagnetic systems and should be related to the particular antiferromagnetic 
ordering of the 122 Fe-pnictides, where both local moment and SDW type orderings coexist   \cite{harriger, zhao, liu}.
We note that $\rho_{xy}$  and  $\rho_{xx}$ are also in anti-correlation  in the cuprate superconductors
(HTSC), so that the tangent of the Hall angle in these materials also decreases when the temperature
increases \cite{ong, grayson}. As for the Fe-pnictides, the existence of AHE in the HTSC is a matter of controversy.
However, the presence of antiferromagnetic correlations is relevant in these two systems.
In the pure and slightly doped Fe-pnictides, long-range SDW stabilizes into an antiferromagnetic ordering
at non-zero temperatures. By using quasi-particle interference imaging,
antiferromagnetic spin fluctuations have already been identified as the principal electron-boson interaction in 
Fe-pnictides \cite{allan}.
In the HTSC, spin fluctuations have been detected by neutron scattering experiments 
and may underly the pseudogap phenomenon \cite{dai}. 
These excitations may be short-ranged and dynamical SDW, but can significantly affect the transport properties
if they last long enough in comparison to the carrier relaxation time.

It is also relevant to observe that the temperature dependence of the Hall angle shown 
in Fig. \ref{rxy_rxx} and described by Eq. (\ref{eq_tanT}) extrapolates to a non-zero and large
value at $T = 0$. Within the interpretation that ascribes a magnetic origin for
$\rho_{xy}$  at low temperatures, this result also suggests that the role of spin chiralities
is dominant to explain the Hall effect in the 122 Fe-pnictides.
%  It also interesting to observe 
% in Table \ref{Tan_parameters} that the values for parameters $\alpha$ and $\beta$ are largely
% changed in the doped samples with respect to the pure one; however these parameters are almost the same
% when the impurity atoms are Mn or Cu, but much smaller when the impurity atom is Ni. 
% Taking into account that among the impurity atoms studied here only Ni-doped samples stabilize the superconductor state,
% it would be interesting
% to broad this study to samples with  impurity atoms as Co, Rh or Pd  which in some larger
% concentrations also present a superconducting phase. If the $\alpha$ and $\beta$ parameters 
% are similar in those samples to those obtained here for our ``Ni-doped sample'', it could serve as an indicator 
% if the strength of the magnetic excitations giving place to electron scattering
% is in close relation with the formation of Cooper pairs.

As a final remark, we stress that the multi-band character of the charge carriers in the studied compounds must also 
be taken into consideration in order to completely describe their magneto-transport properties.
Two-band conduction explains the small magnetoresistance and Hall resistivity generally
observed in the paramagnetic high temperature phase. 
Although authors in Ref. \cite{ishida2} consider that three bands, two electron-type and one hole-type, must be
considered to fully account for the magneto-transport properties of the pure BaFe$_2$As$_2$ compound, in the phenomenological
description of the magnetoresistance and Hall effect results in our substituted compounds, it seems good enough to consider a 
simpler picture where a single current of electron-type carriers adds to conduction by holes. 
Our results indicate that multi-band conduction alone hardly explains the 
behavior of both  properties in the low-temperature ordered phase, 
since one has to suppose that besides the occurrence of a  drastic Fermi surface reconstruction  at $T_{\text{SDW}}$,
progressive modifications of the band structure ocurr below this temperature.
A physically more simple description of the magneto-transport properties of the 122 Fe pnictides might be achieved by 
considering the effect of scattering by magnetic excitations, which have been proven to be important in these
compounds \cite{fernadesv, borisenko, allan}.

% In addition, one is faced to unlikely physical parameters when trying to conciliate the magnetic field and
% temperature dependencies of the magnetoresistance amplitude and the Hall coefficient in the 
% temperature region below $T_{\text{SDW}}$ within the two-band conduction scenario.

% As a consequence
% of this modification, the electron mobility should become progressively much larger than the hole mobility
% as the temperature decreases below $T_{\text{SDW}}$ . According to the temperature dependence of the magnetoresistance 
% amplitude shown in Fig. \ref{MR_T}, the difference between these mobilities should saturate as the 
%temperature approaches zero. 
% This behavior is difficult to conciliate with the temperature dependence of the Hall coefficient
% within the two-band approach. This quantity also depends on the difference between electron and hole mobilities,
% but continuously increases as the temperature decreases toward zero.  Still more difficult to explain within a 
% simple two-band conduction picture is the behavior of the tangent of the Hall angle that extrapolates to a non-zero 
% value at T = 0. All those results indicate that magnetic excitations must be taken into account, and are probably 
%dominant,
% to explain the magneto-transport properties of the 122 Fe-pnictides in the temperature range below the N\'eel
%temperature.     

\section{Summary}

We have measured  the electrical resistivity, magnetoresistance and Hall effect in 
slightly substituted samples of the  BaFe$_{2-x}$TM$_x$As$_2$ system. Four single crystals were investigated,
one pure sample with no Fe substitution, and three samples where TM = Mn, Cu, and Ni. 
Within the studied concentration regime, all three types of substituting atoms diminish the magnetic
 temperature transition by approximately 15 K with respect to that of the pure sample.
From our resistivity and magneto-transport
experiments we conclude that in the low dilution limit the 
chemical substitutions at the Fe sites rather behave  as scattering centers, and
 little changes are produced in the carrier density.
%  and the overall behavior of the magneto-transport properties; in general
% we conclude  that the physical nature of the stabilized  magnetic phase do not depend much on the chemical nature 
% of the impurity atom.

The obtained results were discussed with basis on
a scenario where the magneto-transport phenomena in the magnetically ordered phase of 122 Fe-pnictides 
are mostly governed by magnetic excitations. 
We then assume the occurrence of magnetic, or anomalous, 
contributions to the magnetoresistance and to the Hall effect in the ordered state. Indications of 
the validity of our approach are: (i) the temperature dependence of the magnetoresistance amplitude is reminiscent
of a magnetic order parameter which becomes non-zero at $T \leq T_{\text{SDW}}$; (ii) a large and temperature
dependent anomalous Hall contribution explains the enormous increase in the absolute value of the total Hall coefficient 
in the magnetic phase; (iii) the anomalous Hall resistivity does not show a simple power
law scaling with the longitudinal resistivity; (iv) the tangent of the Hall angle varies linearly
with the temperature below  $T_{\text{SDW}}$  and 
extrapolates to a finite value at $T = 0$,suggesting that the anomalous Hall effect of the 122 Fe-pnictides
is due to the spin-chirality mechanism.

Finally, the multi-orbital character of the charge carriers and some Fermi surface reconstruction
at the  structural and magnetic transition might be taken into account
 in a detailed description of the magneto-transport properties of the studied compounds. 
In particular, two-band conduction is the most natural explanation for the practically
zero magnetoresistance and  the very small Hall resistivity that are systematically observed in the paramagnetic high temperature phase.
 In the magnetic ordered phase, however, scattering by magnetic excitations plays a relevant role.
% Entirely rigorous calculations should take into account the absolute and relative variations
% on the conductivities of the charge carriers, the weak temperature dependence of the magnetic susceptibility
% and the anomalous contribution from the scattering produced because of magnetic ordering.

\section*{Acknowledgments}

This work was partially financed by the Brazilian agencies FAPERGS, FAPESP (Grants  2012/05903-6, 2012/04870-7 and \mbox{2011/01564-0}),
and CNPq (Grants PRONEX 10/0009-2, 304649/2013-9 and 442230/2014-1).
J.P. Pe\~na benefits from a CNPq fellowship.

\bibliographystyle{abbrv}

\begin{thebibliography}{}
\bibitem{hosono} Y. Kamihara, T. Watanabe, M. Hirano, and H. Hosono, J. AM. CHEM. SOC. 130 (2008) , 3296-3297.
\bibitem{paglione} J. Paglione and R. Greene, Nature Physics 6 (2010) 645-658.
\bibitem{singh}  D. J. Singh, Physica C 469 (2009) 418-424.
\bibitem{dong} J. Dong \textsl{et. al.}, EPL 83 (2008) 27006. 
\bibitem{harriger} L. W. Harriger, H. Q. Luo, M. S. Liu, C. Frost, J. P. Hu, M. R. Norman, and P. Dai, Phys. Rev. B 84 (2011) 054544.
\bibitem{zhao} J. Zhao,  \textsl{et. al.}, Nature Phys. 5 (2009) 555-560.
\bibitem{liu} M. Liu, \textsl{et al.}, Nature Phys. 8 (2012) 376-381.
\bibitem{ishida} K. Ishida, Y. Nakai, and H. Hosono, JPSJ 78 (2009) 062001.
\bibitem{piva} M. M. Piva \textsl{et. al.}, J. Phys. Condens. Matter 27 (2015) 145701.
\bibitem{garitezi} T. M. Garitezi \textsl{et. al.}, Braz J Phys 43 (2013) 223–229.
\bibitem{canfield} P. C. Canfield and S. L. Bud'ko, Annu. Rev. Condens. Matt. Phys. (2010)  27-50.
\bibitem{ni} N. Ni, A. Thaler, A. Kracher, J. Q. Yan, L. Bud'ko, and P. C. Canfield, Phys Rev. B 80 (2009) 024511.
\bibitem{yamazaki} T. Yamazaki, \textsl{et. al}., Phys. Rev. B 81 (2010) 224511.
\bibitem{rosa} P. F. S. Rosa \textsl{et. al.}, Scientific Reports 4 (2014) 6252.
\bibitem{kim} M. G. Kim, \textsl{et. al}, Phys. Rev. B 82 (2010) 220503.
\bibitem{sefat} A. S. Sefat, \textsl{et. al}., Phys. Rev. B 79 (2009) 224524.
\bibitem{kuo} Hsueh-Hui Kuo \textsl{et. al.}, Phys. Rev. B 84 (2011) 054540. 
\bibitem{huynh} K. K. Huynh, Y. Tanabe, and K. Tanigaki, Phys. Rev. Lett. 106  (2011) 217004.
\bibitem{albenque2}F. R. Albenque, D. Colson and A. Forget, Phys. Rev. B. 88 (2013) 045105.
\bibitem{albenque} F. R. Albenque, D. Colson, A. Forget and H. Alloul, Phys. Rev. Lett. 103  (2009) 057001.
\bibitem{fang} L. fang  \textsl{et. al.}, Phys Rev. B 80 (2009) 140508(R).
\bibitem{yi} M. Yi, \textsl{et al.}, Phys. Rev. B 80 (2009) 174510. 
\bibitem{chu} Jiun-Haw Chu, and J. G. Analytis, and C. Kucharczyk, and I. R. Fisher, Phys. Rev. B 79  (2009) 014506.
\bibitem{olariu} A. Olariu, F. Rullier-Albenque, D. Colson, and A. Forget, Phys. Rev. B 83 (2011) 054518.
\bibitem{meaden} G. T. Meaden, Electrical Resistivity of Metals, Heywood Books, London, UK, 1966.

\bibitem{wang} Y. Wang \textsl{et. al.}, Phys. Rev. Lett. 114 (2015) 097003.

\bibitem{ziman} J. M. Ziman, Principles of the Theory of Solids,  2nd Edition, Cambridge University Press,  1972.

\bibitem{kasuya} T. Kasuya, Progress of Theoretical Physics, vol. 16(1) (1956) 1.

\bibitem{mackenzie} R. H. Mackenzie, J. S. Qualls, S. Y. Han and J. S. Brooks, Phys. Rev. B 57 (1997) 11854.
% \bibitem{gosh} N. Ghosh \textsl{et. al.}, Solid State Communications 150(39–40), 1940 (2010)
% \bibitem{chan} M. K. Chan, \textsl{et. al.}, Phys. Rev. Lett. 113, 177005 (2014)

\bibitem{garitezi2} T. M. Garitezi \textsl{et. al.}, J. Appl. Phys. 115 (2014) 17D711.
\bibitem{rosa2} P. F. S. Rosa, C. Adriano, T. M. Garitezi, T. Grant, Z. Fisk, R. R. Urbano and P. G. Pagliuso, Scientific Reports 4 (2014) 6543.
\bibitem{yin2} Z. P. Yin, K. Haule and G. Kotliar, Nat. Mater. 10 (2011)  932.

\bibitem{fanfarillo} L. Fanfarillo, E. Cappelluti, C. Castellani and L. Benfatto, Phys. Rev. Lett. 109 (2012) 096402.
\bibitem{breitkreiz1} M. Breitkreiz. P. M. R. Brydon, and C. Timm, Phys. Rev. B 88 (2013) 085103.
\bibitem{breitkreiz2} M. Breitkreiz. P. M. R. Brydon, and C. Timm, Phys. Rev. B 89 (2014)  245106.
\bibitem{yin} Z. P. Yin, K. Haule and G. Kotliar, Nature Phys. 7 (2011) 294.




% \bibitem{richard} P. Richard, K. Nakayama, T. Sato, M. Neupane, Y.M. Xu, J. H. Bowen, G.F. Chen, J.L. Luo, N.L. Wang,
% X. Dai, Z. Fang, H. Ding, and T. Takahashi, PRL 104, 137001 (2010)
% \bibitem{ran} Y. Ran, F. Wang, H. Zhai, A. Vishwanath, and D.H. Lee, Phys. Rev. B 79,014505 (2009)
% \bibitem{harrison} N. Harrison,  and S.E. Sebastian, Phys. Rev. B 80, 224512 (2009)
\bibitem{sinitsyn} N. A. Sinitsyn,  J. Phys. Condens. Matter 20 (2008) 023201.
\bibitem{kubler} J. Kubler and C. Felser, EPL 108 (2014) 67001.
\bibitem{chen} H. Chem, Q. Niu and A. H. Mac Donald, Phys. Rev. Lett.112 (2014) 017205.
\bibitem{chendong} X. Chen, S. Dong, and J.M. Liu, Phys. Rev. B 81 (2010) 064420.
\bibitem{tran} V. H. Tran, Mater Sci-Poland, vol. 26(4) (2008)
\bibitem{jolanta} J. Stankiewicz, A. D. Bianchi and Z. Fisk, JPCS 200 (2010) 012192.

\bibitem{fernadesv} R. M. Fernandes, and O. Vafek, Phys. Rev. B 90 (2014) 214514.
\bibitem{borisenko} S. V. Borisenko \textsl{et. al.}, Arxiv:1409.8669v2 (2015)
\bibitem{rodrigues} J. P. Rodrigues, and E. H. Rezayi, Phys. Rev. Lett. 103 (2009) 097204.
\bibitem{nagaosa} N. Nagaosa, J. Sinova, S. Onoda, A. H. McDonald, and N.P. Ong, Rev Mod. Phys. 82 (2010) 1539.
\bibitem{karpus}   R. Karplus and J. M. Luttinger, Phys. Rev. 95 (1954) 1154.
\bibitem{taguchi} Y. Taguchi \textsl{et. al.}, Science 291  (2001) 2573.
\bibitem{thaler} A. Thaler \textsl{et. al.}, Phys. Rev. B 84 (2011)144528
\bibitem{sefat2} A. S. Sefat, R. Jin, M. A. McGuire, B. C. Sales, D. J. Singh, and D. Mandrus, Phys. Rev. Lett. 101 (2008) 117004
\bibitem{kawamura} H. Kawamura, Phys. Rev. Lett. 90 (2003) 047202. 
\bibitem{tatara} G. Tatara and H. Kohno, Phys. Rev. B 67 (2003)  113316.
\bibitem{wolff} F. Wolff-Fabris, P. Pureur, J. Schaf, V. N. Vieira, and I. A. Campbell,
Phys. Rev. B 74 (2006) 214201.
\bibitem{ong} N. P. Ong, Physical Properties of the High Temperature Superconductors, Vol 2,
World Scientific Singapore, 1990.
\bibitem{grayson} M. Grayson, L. Rigal, D. C. Schmadel, H. D. Drew and P.J. Kung  Int. J. Mod. Phys. B 16 (2002) 3148.
\bibitem{allan} M. P. Allan \textsl{et. al.}, Nature Physics vol. 11 (2015) 177. 
\bibitem{dai} P. Dai \textsl{et. al.}, Science 284 (1999) 1344.

\bibitem{ishida2} S. Ishida \textsl{et. al.}, Phys. Rev. B 84 (2011) 184514.

% \bibitem{kuroki} K. Kuroki,  H. Usui, S. Onari,  R. Arita, and H.  Aoki, Phys. Rev. B 79, 224511 (2009)
% \bibitem{ning} Ning, F. L. \textsl{et. al.}, Phys. Rev. Lett. 104, 037001 (2010)
% %\bibitem{fernandes} R.M. Fernandes, A.V. Chubukov, and J. Schmalian, Nature Physics 10, 97-104 (2014)
% %\bibitem{rosental} E. P. Rosenthal, et. al., Nature Phys. 10, 225–232 (2014)
% %\bibitem{johnson} P. D. Johnson, H.B. Yang, J.D. Rameau, G.D. Gu, Z.H. Pan, T. Valla, M. Weinert, and A. V. Fedorov, Phys. Rev. Lett. 114, 167001 (2015)
% \bibitem{sundaram} G. sundaram and Q. Niu, Phys Rev. B 59, 14915 (1999)
% \bibitem{schoenes} J. Schoenes and J.J.M. Franse, Phys. Rev. B 33, 5138(R) (1986)
% \bibitem{fert1} A. Fert,  J. Phys. C (Sol. St. Phys.), 1784–1788 (1969)
% \bibitem{fert2} A. Fert and I. A. Campbell, J. Phys. F: Metal Phys., 849–870 (1976)
% \bibitem{calderon} M.J. Calderon, G. Leon, B. Valenzuela and E. Bascones, Phys. Rev. B 86, 104514 (2012) 
% %\bibitem{alok}  http://meetings.aps.org/link/BAPS.2014.MAR.T13.9
% %\bibitem{sacramento} P.D. Sacramento, M.A.N. Araujo, V.R. Vieira, V.K. Dugaev and J. Barnas, Phys. Rev. B 85, 014518 (2012)
% \bibitem{quin} M.H. Quin, S. Dong, H.B. Zhao, Y. Wang, J.M. Liu and Zhifeng Ren, New Journal of Physics 16, 053027 (2014)
% \bibitem{wiesenmayer} E. Wiesenmayer \textsl{et. al.}, Phys. Rev. Lett. 107, 237001 (2011)
% \bibitem{reisinger} D. Reisinger, P. Majewski, M. Opel, L. Alff, and R. Gross, App. Phys. Lett. vol. 85(21), 4980 (2004)
% \bibitem{westerburg} W. Westerburg, D. Reisinger, and G. Jakob, Phys. Rev. B 62, R767 (2000)


\end{thebibliography}

\end{document}